\documentstyle[aps,epsfig,float,amstex,axodraw]{revtex}

\newcommand{\be}{\begin{equation}}
\newcommand{\ee}{\end{equation}}
\newcommand{\beq}{\begin{eqnarray}}
\newcommand{\eeq}{\end{eqnarray}}
\newcommand{\bi}{\bibitem} 

\restylefloat{figure}
\restylefloat{table}

\begin{document}
    
\def\gC{\mbox{\boldmath $C$}}
\def\gZ{\mbox{\boldmath $Z$}}
\def\gR{\mbox{\boldmath $R$}}
\def\gN{\mbox{\boldmath $N$}}
\def\ua{\uparrow}
\def\da{\downarrow}
\def\eq{\equiv}
\def\a{\alpha}
\def\b{\beta}
\def\g{\gamma}
\def\G{\Gamma}
\def\d{\delta}
\def\D{\Delta}
\def\e{\epsilon}
\def\z{\zeta}
\def\h{\eta}
\def\th{\theta}
\def\k{\kappa}
\def\l{\lambda}
\def\L{\Lambda}
\def\m{\mu}
\def\n{\nu}
\def\x{\xi}
\def\X{\Xi}
\def\p{\pi}
\def\P{\Pi}
\def\r{\rho}
\def\s{\sigma}
\def\S{\Sigma}
\def\t{\tau}
\def\f{\phi}
\def\vf{\varphi}
\def\F{\Phi}
\def\c{\chi}
\def\w{\omega}
\def\W{\Omega}
\def\Q{\Psi}
\def\q{\psi}
\def\de{\partial}
\def\inf{\infty}
\def\ra{\rightarrow}
\def\bra{\langle}
\def\ket{\rangle}

\title{The ``$W=0$'' Pairing Mechanism from Repulsive Interactions in 
Symmetric $2d$  Models}

\author{Michele Cini and Gianluca Stefanucci}

\address{Istituto Nazionale di Fisica della Materia, Dipartimento di Fisica,\\
Universita' di Roma Tor Vergata, Via della Ricerca Scientifica, 1-00133\\
Roma, Italy}
\maketitle

\begin{abstract}     

{\small
In two-dimensional systems possessing a high degree of symmetry, the  repulsive
electron-electron interaction produces a pairing force; the mechanism  
would fail  in the presence of strong 
distortions. We have studied this in the one-band and three-band  
Hubbard Model. 
From partially occupied orbitals one obtains pair eigenstates
of the Hamiltonian with no on-site repulsion (the $W=0$ pairs).
The concept of $W=0$ pairs allows to make qualitative and quantitative
predictions about the behaviour of 
interacting many-body systems, a quite remarkable and 
unusual situation.
Exact numerical solutions  for clusters with ``magic'' hole 
numbers reveal attraction between the holes in $W=0$ pairs. The 
effect occurs in all fully symmetric clusters which    are  centered on a Cu site;
then holes get paired in a 
wide, physically relevant parameter range and show superconducting 
quantization of the magnetic flux. A canonical transformation of the Hamiltonian, valid for 
clusters and for the full plane, leads to a Cooper-like equation for 
the $W=0$ pairs. We have evaluated the effective interaction and 
found that $W=0$ pairs are 
the bare quasiparticles which, once dressed, become two-hole bound 
states. We applied the above theory to the doped 
antiferromagnet, and found that the ground state at half filling is the 
singlet component of a determinantal state. We  
write down this determinant and the ground state wave function explicitly in terms of a
many-body $W=0$ eigenstate. 
Our analytical results for the $4\times 4$ square lattice at half 
filling and with doped holes, compared to available numerical data,
demonstrate that the method, besides providing intuitive grasp on 
pairing, also has quantitative  predictive power. }
\end{abstract}

\section{Introduction}
\label{intro3}

{\small
Most superconducting oxides\cite{bedmu} have large unit cells containing four or 
more elements and CuO$_{2}$ planes separated by intermediate layers 
which stabilize the lattice and act as a charge reservoir. A large 
amount of experimental and theoretical work has clarified that mobile 
holes on these planes are responsible for superconductivity. In the 
undoped solid the electrons are so strongly correlated that they 
cannot move freely as in a metal (antiferromagnetic Mott insulator). 
When holes are introduced into the 
plane by doping, the system becomes conducting and even 
superconducting provided the hole density is within the appropriate 
range. Increasing experimental evidence obtained in several cuprate superconductors
has convincingly demonstrated that the pairing state of these materials has 
a non-trivial symmetry, ranging from $s$- to $d$-wave\cite{harlingen}. 
Moreover, the pairs exist above the critical
temperature either in the form of superconducting fluctuations or preformed
pairs. The latter aspect is apparent in the underdoped region in
which a clear pseudogap essentially of the same magnitude as the
superconducting gap is measured in ARPES\cite{marshall}, 
NMR\cite{berthier}, neutron scattering\cite{rossat} and specific heat 
measurement\cite{loram} in metallic 
phase. All these signatures put strict
constraints to any microscopic model of the cuprates. Any theory of the
paired state must predict the correct symmetry and doping dependence of the
binding energy of the pair, but the pairing mechanism must be
doping-independent and robust enough to survive superconducting fluctuations
well into the normal state, far from optimum doping.

The class of theories based on correlation 
mechanisms start from single or multiple bands Hubbard-like 
Hamiltonians containing nearest neighbors hopping matrix elements $t$ 
between Cu and O and O-O and large on-site Coulomb repulsive energies 
$U$ between holes of opposite spins which tends to reduce double 
occupancy of these sites. For $U>>t$, such Hamiltonians map into 
Heisenberg-like Hamiltonians for holes propagating through an 
antiferromagnetic background with an exchange interaction $J$ between 
neighboring spins. Perturbative approaches such 
as the t-J Model\cite{dagrev}\cite{sush}, by means of the Guzwiller 
projection\cite{gutz}  
force the holes to avoid on site interactions by 
letting $U$ to be very large (or infinite) so that eventually the 
pairing may be achieved through residual small attractive 
interactions. 
In this way the problem is circumvented formally, but not understood 
physically. Such 
models do not identify from the outset the physical mechanism capable 
to nullify the large repulsive Coulomb interactions among the paired holes.
On one hand, the Guzwiller projection cuts a large part of the Hilbert 
space, and that would cost a catastrophic amount in kinetic energy, 
compared to the energy gain involved in the normal to superconductor 
transition, unless the 
repulsion is orders of magnitude larger than the hopping, which is {\em not} 
the case; on the other hand, the same idea could equally well be applied to a potato, 
i.e. it does not explain what is special in the CuO$_{2}$ plane.
 
Due to the small size of the Cooper pairs, we believe that any serious 
proposal of a pairing mechanism should essentially be size-independent, 
i.e. it should work in the thermodynamic limit but it should 
be manifest in tiny systems as well. Exact calculations on finite
models should bring to light interesting local
aspects of the microscopic pairing mechanism, and  be useful as tests for the
analytic developments; pairing should be possible using a physically 
plausible range of  parameter values. 

In this review article, we illustrate the ``$W=0$'' pairing mechanism, which is 
specific of the square $2d$ structure of the  CuO$_{2}$ plane; instead of 
forcing the system by the Guzwiller projection we get the attraction by using the 
symmetry. The symmetry provides a physical mechanism capable of cutting down the 
on-site repulsion from the outset.  We want to show that the holes avoid 
the on site interaction via a correlation mechanism which is 
effective for any $U/t$. We demonstrate this correlation effect, 
showing that its basic ingredient is a configuration mixing driven 
by the symmetry of the system which entails the presence of degenerate 
hole states at the Fermi level. Moreover, we provide 
evidence that the same configuration mixing, 
when applied to the many-body problem, also 
produces pairing 
between the holes and superconducting flux quantization  
when physical parameter values are 
used. 

The work is divided into two main parts. In the first part, that is 
Section \ref{chapter3}, we 
 review the pairing mechanism in small symmetric clusters. 
We begin by showing that by symmetry the Hubbard Hamiltonian 
admits two-hole singlet eigenstates 
that  do not feel the on-site Hubbard repulsion, that we call $W=0$ pairs.  
First, we demonstrate that the prototype 
CuO$_{4}$  cluster already contains the essential ingredients of 
the mechanism. Then we prove a general 
theorem on the allowed symmetries of $W=0$ pairs. From the theorem we 
extract a practical technique to build them explicitly for symmetric 
clusters and for the full plane. 
We will see that no $W=0$ pairs exist if the symmetry is too low. 
In Section \ref{ciandexsol} we show that the peculiar properties 
of $W=0$ pairs cause a stabilization of the interacting system which 
leads to pairing. A careful analysis is made 
on the smallest ``allowed'' cluster, the CuO$_{4}$; then, the 
investigation is 
extended to larger fully symmetric clusters. It is shown that 
pairing can be obtained in a physical parameter range 
any time the number of holes is ``magic''. Section \ref{effecinter} 
is devoted to the analytic interpretation of the results. It is shown 
that in second order perturbation theory the effective interaction 
between the holes of a $W=0$ pair comes out from the interference of 
positive and negative contributions, depending on the point symmetry 
of the electron-hole states exchanged, and that it may be attractive. 
Finally in Section \ref{fluxqinsymclusters} we demonstrate  the 
supercounducting flux quantization and conclude that $W=0$ pairs are 
the bare quasiparticles which become two-hole bound states when 
dressed by the interaction.

In the second part, that is Section \ref{chapter4}, 
we generalize the theory to the full plane. 
In Section \ref{bands}, we review the band structure of the 
three-band Hubbard Model and fix some notations. In Section 
\ref{w=0andflux} we show how $W=0$ pairs in the full plane arise at the Fermi 
level for any filling, and that their symmetry properties naturally 
lead to superconducting flux-quantisation. Section \ref{canonical} 
is devoted to the general canonical transformation leading to the
effective Hamiltonian for pairs. The method is new and free from the
limitations of perturbation theory; the relation of the present formalism 
to Cooper theory from one side and to the cluster results from the other is 
discussed. The final result of the analysis is an integral equation 
for pairs in the full plane, which is suitable for a numerical study.  
Therefore, in Section \ref{numerical} the integral equation is solved 
numerically.  We first solve it in finite supercells and then take the
asymptotic limit. Two kinds of bound states of different symmetries result, and 
the dependence of $\Delta$ on the filling and other parameters is 
explored. Next, we study the Hubbard Model at and near half filling. Here the 
problem is that the non-interacting ground state at half filling is 
highly degenerate, while the canonical transformation needs a unique 
non-interacting ground state to work. Therefore, in Section \ref{afgs} 
we remove the degeneracy in first order perturbation theory by means 
of a suitable formalism, the {\it local} formalism, together with the 
help of a theorem  by Lieb. The half filled ground state at weak 
coupling is explicitly written down. We show that it is the 
spin singlet projection of a peculiar determinantal state; the latter 
which exhibits the {\it antiferromagnetic 
property}: the translation by a lattice step is equivalent to a spin flip. 
In Section \ref{dopedaf} we use the $4\times 4$ square lattice as a test 
case. The half filled antiferromagnetic ground state is doped with two holes and 
an effective interaction between them is derived by means of the 
canonical transformation. The analytical results agree well with the 
numerical ones and this shows the predictive power of the approach.  
Finally, the main conclusions are summarized in Section 
\ref{conclusions}.}

\section{Pairing in the Hubbard Model: Symmetric Clusters} 
\label{chapter3}

{\small 
Our starting point is the three-band Hubbard Model 
\begin{equation}
H=K+W_{\mathrm{Hubbard}}+W_{\mathrm{off-site}}
\label{tbhh}
\end{equation}
where
\begin{equation}
K=\sum_{ss',\s}t_{ss'}c^{\dag}_{s\s}c_{s'\s}=
t_{pd}\sum_{\bra ij\ket,\s}(c^{\dag}_{j\s}c_{i\s}+{\mathrm h.c.})+t_{pp}\sum_{\bra 
jj'\ket,\s}(c^{\dag}_{j\s}c_{j'\s}+{\mathrm h.c.})+\e_{d}\sum_{i,\s}\hat{n}_{i\s}+
\e_{p}\sum_{j,\s}
\hat{n}_{j\s}
\end{equation}
and
\begin{equation}
W_{\mathrm{Hubbard}}=U_{d}\sum_{i}\hat{n}_{i\ua}\hat{n}_{i\da}+U_{p}
\sum_{j}\hat{n}_{j\ua}\hat{n}_{j\da},\;\;\;\;\;
W_{\mathrm{off-site}}=U_{pd}\sum_{\bra ij\ket,\s\s'}\hat{n}_{i\s}\hat{n}_{j\s'}.
\end{equation}
Here, $c_{j}$ ($c_{i}$) are fermionic operators that destroy holes at the Oxygen 
(Copper) ions 
labelled $j$ ($i$) and $\hat{n}=c^{\dag}c$ is the number operator. 
$\bra ij\ket$ refers to pairs of nearest 
neighbors $i$ (Copper) and $j$ (Oxygen) sites. The hopping terms  
corresponds to the hybridization between nearest neighbors Cu and O 
atoms, and are roughly proportional to the overlap between localized 
Wannier orbitals.  $U_{d}$ 
and $U_{p}$ are positive constants that represent the repulsion 
between holes when they are at the same Copper ($d$) and Oxygen ($p$) orbitals, 
respectively. $U_{pd}$ has a similar meaning, {\it i.e.} it 
corresponds to the Coulombic repulsion when two holes occupy two 
adiacent Cu-O sites.  The on site energies $\e_{p}$ and 
$\e_{d}$  represent the 
difference in energy between the occupied orbitals of Oxygen and 
Copper. In the strong coupling limit, and with one particle per unit 
cell, this model reduces to the spin Heisenberg Model with a 
superexchange antiferromagnetic coupling\cite{emery2}\cite{fulde}.

The Hamiltonian in Eq.(\ref{tbhh}) shows that for 
$\D_{pd}=\e_{p}-\e_{d}>0$, the first hole added to the system will 
energetically prefer to occupy mostly the $d$-orbital of the Copper 
ions. This is indeed the observed situation in 
the undoped materials which have one hole per unit cell. When another 
hole is added to this unit cell, and working in the regime where 
$U_{d}$ is larger than $\D_{pd}$, the new hole mainly occupy Oxygen 
orbitals\cite{fink}. 
This is in agreement with Electron Energy Loss Spectroscopy 
experiments\cite{nucker}. From a band structure and by best fitting 
the results of up-to-date {\it ab initio} 
calculations\cite{hyber} one can roughly estimate the actual values 
of the parameters in the Hamiltonian of Eq.(\ref{tbhh}). Our preferred 
set\footnote{We are using 
a symmetric gauge; for instance, F. C. Zhang 
and T. M. Rice, {\it Phys. Rev.} {\bf B 37}, 3759 (1988) use a different 
sign prescription for $t_{pd}$ and $t_{pp}$, which corresponds to 
changing the sign of some orbitals; using their convention, the orbital 
symmetry labels $a_{1}$ and $b_{1}$ get exchanged.} is 
shown in Table \ref{parameters} (in eV). 
\begin{table}[tbp]
    \centering
    \caption{{\footnotesize Hole parameters (in eV) for the 
    three-band Hubbard Model in Eq.(\ref{tbhh}).}}
    \vspace{0.5cm}
    \begin{tabular}{|c|c|c|c|c|c|}
        \hline
        $\e_{p}-\e_{d}$ & $t_{pd}$ & $t_{pp}$ & $U_{d}$ & 
        $U_{p}$ & $U_{pd}$  \\
        \hline
        3.5 & 1.3 & -0.65 & 5.3 & 6 & $<$ 1.2  \\
        \hline
    \end{tabular}
    \label{parameters}
\end{table}

Several years ago, Hirsch and Scalapino\cite{scalah} introduced the quantity 
$\D(N)=E_{0}(N)+E_{0}(N-2)-2E_{0}(N-1)$, where $E_{0}(N)$ is the ground state energy 
of the system with $N$ holes. They showed by direct diagonalization of 
several clusters that $\D$ can be negative with all positive Coulomb 
repulsion parameter. The $\D<0$ behavior was taken as an {\em indication} 
that some kind of attraction was developing from repulsive interaction. 
From those 
calculations\cite{alascio}, it was apparent that negative $\D$ arose 
from unphysically large Cu-O off-site interactions $U_{pd}$ of the order of 4-6 eV at 
least. Moreover the $\D<0$ 
criterion has also been questioned\cite{mazu} because when one introduces the
nuclear degrees of freedom, the sign of $\D$ can be reversed. This can
happen when the ground state with $N-1$ holes is degenerate, and Jahn-Teller
distorts, gaining energy by the distortion; the pairing then looks like an
artifact due to the neglect of vibrations. Ultimately, the argument runs
into trouble because it depends on a comparison of ground state energies
with different hole numbers {\it N}.

In 1995, Cini and Balzarotti\cite{cb1} 
resumed the subject from a new standpoint which avoids the 
need for a large $U_{pd}$ and is free from the above criticisms.  
They started from the observation that 
highly-symmetric clusters possess 2-holes singlet eigenstates which do not 
feel the on-site repulsion $W_{\mathrm{Hubbard}}$, {\it $W=0$ pairs}. 
This property  poses severe limitations on 
finite clusters that not only must possess the full $C_{4v}$ symmetry, 
but must be centered around a Cu site, see Figure \ref{symclusters}. 
\begin{figure}[H]
\begin{center}
	\epsfig{figure=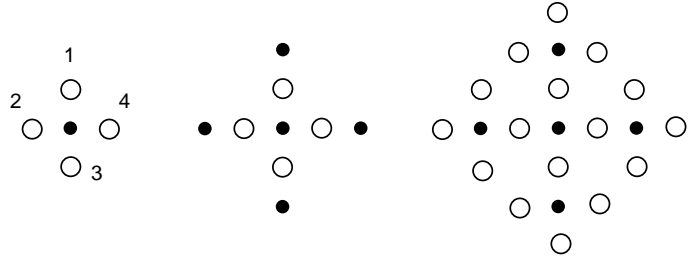,width=9cm}
	\caption{\footnotesize{
	CuO$_{4}$, Cu$_{5}$O$_{4}$ and 
    Cu$_{5}$O$_{16}$ clusters, the ``allowed'' geometries. Full dots 
    are Cu ions.}}
    \label{symclusters}
\end{center} 
\end{figure}

Such $W=0$ eigenstates are 
generally excited states of the two-body system and arise when the 
two holes occupy degenerate states of $E \equiv (x,y)$ symmetry.
The interesting situation arises when the hole population is such that 
the {\em non interacting ground state} would  have two $E$ holes.
Thus, any allowed cluster has its ``magic''  number of holes. The {\em 
interacting ground state} must be described in terms of a many-body
configuration mixing, but the presence of the $W=0$ pair imparts to the 
many-body state a special character, as we shall show; the 
persistence of the ``$W=0$'' character in the interacting case is the 
 main point that we want to make. 
The need for a partially filled shell restricts the hole numbers
for small clusters, whereas of course any even number of holes works in the thermodynamic
regime. In the infinite plane, $W=0$ pairs are easily obtained\cite{cb2} 
from the non-bonding band; a method for building them at any kinetic 
energy is discussed below. 

Consider the process of filling the one-body levels of the 
noninteracting system according to the {\em aufbau} principle.
Since the order of the one-body energy levels depends on the site 
energies and hopping integrals, for a given cluster the ``last pair'' is 
a $W=0$ pair in part of the parameter space. Therefore, $W=0$ pairs 
fix the cluster structure, occupation and the range of the 
parameters. These restrinctions are so stringent that 
these clusters with such properties had not been studied 
previously. In particular, the present discussion does not apply to 
the forbidden geometries, like those examined by Hirsch {\it et 
al.}\cite{scalah}\cite{scalah2} and Balseiro {\it et 
al.}\cite{alascio}. The Cu$_{4}$O$_{4}$ geometry considered by Ogata 
and Shiba\cite{ogata} has the $C_{4v}$ symmetry of the lattice, but 
lacks the central Cu, and therefore it is forbidden (the pairs on 
degenerate levels feel the on-site repulsion, as one can see by 
performing the two-hole configuration interaction 
calculation\cite{cb1}). The dynamics of holes in Cu$_{4}$O$_{4}$ does 
not lead to pairing if the on-site repulsion on Oxygen is 
included\cite{martin}. }

\subsection{$W=0$ Pairs}
\label{w=opairs}

{\small
In this Section, we limit the discussion to on-site interactions 
($U_{pd}=0$), while the effect of the off-site terms is deferred to 
Section \ref{ciandexsol}.}

\subsubsection{$W=0$ Pairs in the CuO$_{4}$ Planar Cluster}
\label{w=opairscuo4}

{\small 
For the sake of simplicity, let us first consider the 
CuO$_{4}$ planar cluster of Figure \ref{symclusters} with 4 holes, 
one intrinsic and three due to doping. Site 1 is Cu, and the others 
are Oxygens. The single-hole energy level scheme {\it vs.} $t_{pp}$ 
is shown in Figure \ref{1holespectrumcuo4}.
\begin{figure}[tbp]
\begin{center}
	\epsfig{figure=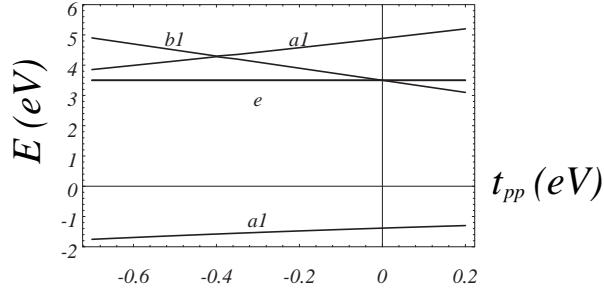,width=8cm}
	\caption{\footnotesize{
	One-hole energy levels of the CuO$_{4}$ 
    cluster {\it vs.} $t_{pp}$; the other parameter are specified in 
    Table \ref{parameters}. The levels $e$ and $b_{1}$ 
    are non-bonding, the $a_{1}$ levels form a bonding-antibonding pair. 
    The levels are labelled according to the irreducible 
    representations of the $C_{4v}$ symmetry Group.}}
    \label{1holespectrumcuo4}
\end{center} 
\end{figure}
The levels are labelled 
according to the irreducible representations of the $C_{4v}$ symmetry 
Group whose character table is shown in Table \ref{c4vtable}.

At $t_{pp}=-0.65$ eV, the first two holes go into a bonding level of 
$A_{1}$ symmetry, and the next two go into a non-bonding level of $E$ 
symmetry, with orbitals transforming like the $(x,y)$ vector. Denoting 
the bonding state by $a$ and the degenerate non-bonding states by 
$e_{x}$ and $e_{y}$, one could build the degenerate singlet 
configurations $|a_{\ua}a_{\da}e_{x\ua}e_{x\da}\ket$ and 
$|a_{\ua}a_{\da}e_{y\ua}e_{y\da}\ket$ and the triplet components 
$|a_{\ua}a_{\da}e_{x\ua}e_{y\ua}\ket$ and 
$|a_{\ua}a_{\da}e_{x\da}e_{y\da}\ket$. 
Since parallel spin holes are not interacting in the model, 
in the Hartree-Fock (HF) approximation the triplet would have lower 
energy according to the Hund's rule. 

Yet, the singlet energy can be much improved by configuration 
interaction (CI). First, let us replace the $(e_{x},e_{y})$ pair by the 
$(\a,\b)$, where $\a=(e_{x}+e_{y})/\sqrt{2}$ and $\b=(e_{x}-e_{y})/\sqrt{2}$ are 
oriented in the O-O directions and are connected by a $\p/2$ rotation. 
The singlet configurations then 
read $|\Q_{\a\a}\ket=|a_{\ua}a_{\da}\a_{\ua}\a_{\da}\ket$ and 
$|\Q_{\b\b}\ket=|a_{\ua}a_{\da}\b_{\ua}\b_{\da}\ket$, both with HF 
energy
\begin{equation}
\bra\Q_{\a\a}|H|\Q_{\a\a}\ket=\sum_{\s}(K_{a\s}+K_{\a\s})+
\sum_{s}U_{s}n_{\ua}(s)n_{\da}(s),
\end{equation}
where $s$ denotes the generic site and $U_{s}=U_{d}$ if $s$ is a Cu 
site and $U_{s}=U_{p}$ otherwise. The first term on the right hand 
side contains the one-body contributions
\begin{equation}
K_{\vf\s}=\sum_{s,s'}\vf^{\ast}_{\s}(s)t_{ss'}\vf_{\s}(s'),\;\;\;\;\;\;\;
\vf(s)\equiv\bra s|\vf\ket,\;\;\;\vf=a,\;\a
\end{equation}
where $t_{ss'}$ is the hopping matrix element of the three-band  
Hubbard Model with diagonal entries $t_{ss}=\e_{d}$ ($\e_{p}$) if $s$ 
is a Cu (O) site, and $n_{\s}(s)=|a_{\s}(s)|^{2}+|\a_{\s}(s)|^{2}
\equiv n_{a}(s)+n_{\a}(s)$. 
The configurations are mixed by 
$\bra\Q_{\a\a}|H|\Q_{\b\b}\ket=\sum_{s}U_{s}n_{\a}(s)n_{\b}(s)$. 
The lowest eigenvalue corresponding to the choice
\begin{equation}
|\Q^{(-)}\ket=\frac{1}{\sqrt{2}}(|\Q_{\a\a}\ket-|\Q_{\b\b}\ket)
\label{ciwfincuo4}
\end{equation}
is
\begin{equation}
\varepsilon=\bra\Q_{\a\a}|H|\Q_{\a\a}\ket-\bra\Q_{\a\a}|H|\Q_{\b\b}\ket=
\sum_{\s}(K_{a\s}+K_{\a\s}) +\sum_{s}U_{s}[n_{a}^{2}(s)+2n_{a}(s)n_{\a}(s)]+W,\;\;
\end{equation}
where $W=\sum_{s}U_{s}n_{\a}(s)[n_{\a}(s)-n_{\b}(s)]$ 
represents the direct interaction between the non-bonding holes. At 
this point the issue is decided by the peculiar property of the 
system that $n_{\a}(s)$ and $n_{\b}(s)$ {\it are identical, despite the 
fact that the quantum states are distinct and orthogonal}, and 
therefore the interaction between those holes is turned off exactly. 
We emphasized that the existence of degenerate states such that 
$n_{\a}(s)=n_{\b}(s)$ is the only condition. This two-hole singlet eigenstate
with vanishing on site repulsion is a $W=0$ {\it pair}. 
 The above argument 
brings out the essential role of symmetry, and can be extended to 
clusters of arbitrary size. The singlet in Eq.(\ref{ciwfincuo4}) turns 
out to be $^{1}B_{2}$ which transforms as $xy$ and has a totally 
symmetric charge distribution. 
\begin{table}[tbp]
    \centering
    \caption{{\footnotesize Character table of the $C_{4v}$ symmetry Group. 
    Here  $\mathbf{1}$ denotes the identity,  $C_{2}$ the  180 degrees 
    rotation,  $C_{4}^{(+)},\;C_{4}^{(-)}$ the 
counterclockwise and clockwise 90 degrees rotation, $\s_{x},\;\s_{y}$ 
the reflection with respect to the $y=0$ and $x=0$ axis and $\s_{+},\;\s_{-}$ 
the reflection with respect to the 
$x=y$ and $x=-y$ diagonals. The $C_{4v}$ symmetry Group has 4 one-dimensional 
irreducible representations (irreps) $A_{1}$, 
$A_{2}$, $B_{1}$, $B_{2}$ and 1 two-dimensional irrep $E$. In the last 
column it is shown how each of them transforms under $C_{4v}$.}}
    \vspace{0.5cm}
   \begin{tabular}{|c|c|c|c|c|c|c|}
      \hline 
$C_{4v}$ & $\mathbf{1}$ & $C_{2}$ & $C^{(+)}_{4},\,C^{(-)}_{4}$ & 
$\s_{x},\,\s_{y}$ & 
$\s_{+},\,\s_{-}$ & Symmetry \\
\hline 
$A_{1}$ & 1 & 1 & 1 & 1 & 1 & $x^{2}+y^{2}$\\
\hline 
$A_{2}$ & 1 & 1 & 1 & -1 & -1 & $(x/y)-(y/x)$\\
\hline 
$B_{1}$ & 1 & 1 & -1 & 1 & -1 & $x^{2}-y^{2}$\\
\hline 
$B_{2}$ & 1 & 1 & -1 & -1 & 1 & $xy$ \\
\hline 
$E$ & 2 & -2 & 0 & 0 & 0 & $(x,y)$ \\
\hline 
\end{tabular}
    \label{c4vtable}
\end{table}

Since $W=0$ for both $^{1}B_{2}$ and the triplet, two-body 
calculations cannot predict which is the ground state. In Section 
\ref{ciandexsol}  
we shall show that pairing is indeed achieved in part of the parameter 
space and a non-degenerate singlet ground state prevails.}

\subsubsection{$W=0$ Theorem}
\label{w=otheorem}

{\small 
For any finite cluster having the full $C_{4v}$ symmetry and a Cu ion 
at the center, one can obtain $W=0$ pair eigenstates\cite{cb34} even 
if the above procedure is not directly extendible. 

Here we want to point out a  more powerful and elegant criterion to get 
all the $W=0$ pair eigenstates of a given system. 
As we shall see this criterion applies 
to Hubbard Models defined on an arbitrary graph $\L$. Let ${\cal G}_{0}$ be the 
the symmetry Group of the 
non-interacting Hubbard Hamiltonian 
$K=\sum_{xy\in\L,\s}t_{xy}c^{\dag}_{x\s}c_{y\s}$. 
By definition, every one-body eigenstate of $H_{\mathrm{Hubbard}}=K+
W_{\mathrm{Hubbard}}$, $W_{\mathrm{Hubbard}}=\sum_{x\in\L}U_{x}
\hat{n}_{x\ua}\hat{n}_{x\da}$, can be 
classified as belonging to one of the irreducible representations 
(irreps) of ${\cal G}_{0}$. We may 
say that an irrep $\eta$ is represented in the one-body spectrum of 
$H_{\mathrm{Hubbard}}$ if 
at least one of the one-body levels belongs to $\eta$. Let
${\cal E}$ be the set of the irreps of ${\cal G}_{0}$ which are 
represented in the one-body spectrum of $H_{\mathrm{Hubbard}}$. 
Let $|\q\rangle$ 
be a two-body eigenstate of 
the non-interacting Hamiltonian with spin $S^{z}=0$ and $P^{(\eta)}$ the 
projection operator on the irrep $\eta$. We wish to prove the 

\underline{{\it $W=0$ Theorem.}} - 
Any nonvanishing projection of $|\q\rangle$ on an irrep
\underline{not} contained in ${\cal E}$, is an eigenstate of $H_{\mathrm{Hubbard}}$ 
with no double occupancy. The singlet component of this state is a  
$W=0$ pair. Conversely, any pair belonging to an irrep represented in the 
spectrum must have non-vanishing $W_{\mathrm{Hubbard}}$ expectation value:
\begin{equation}
\eta \notin  {\cal E} \Leftrightarrow W_{\mathrm{Hubbard}} P^{(\eta)}|\q\rangle=0 
\label{theo}
\end{equation}

{\it Proof}: Let us consider a  
two-body state of opposite spins belonging to the irrep $\eta$ of ${\cal G}_{0}$:
\begin{equation}
|\q^{(\eta)}\rangle=\sum_{xy\in\L}\q^{(\eta)}(x,y)
    c^{\dag}_{x\ua}c^{\dag}_{y\da}|0\rangle.
\end{equation}
Then we have
\begin{equation}
 \hat{n}_{x\ua}\hat{n}_{x\da}|\q^{(\eta)}\rangle=
 \q^{(\eta)}(x,x)c^{\dag}_{x\ua}c^{\dag}_{x\da}|0\rangle\equiv 
    \q^{(\eta)}(x,x)|x\ua,x\da\rangle.
\end{equation}
We define $P^{(\eta)}$ as the projection operator on the irrep $\eta$. Since
\begin{equation}
   P^{(\eta)}\sum_{x\in\L}\q^{(\eta)}(x,x)|x\ua,x\da\rangle=
   \sum_{x\in\L}\q^{(\eta)}(x,x)|x\ua,x\da\rangle
\end{equation}
if $P^{(\eta)}|x\ua,x\da\rangle=0\;\forall x\in\L$ 
then $\q^{(\eta)}(x,x)=0\;\forall x\in\L$. 
It is worth to note that this condition is satisfied if and only if 
$P^{(\eta)}|x\s\rangle=0\;\forall x\in\L$ 
where $|x\s\rangle=c^{\dag}_{x\s}|0\ket$. 

Now it is always possible to 
write $ |x\s\rangle$ in the form  $|x\s\rangle=\sum_{\eta\in {\cal E}}c^{(\eta)}(x)
|\eta_{\s}\rangle$ where $|\eta_{\s}\rangle$ is the one-body eigenstate of 
$H_{\mathrm{Hubbard}}$ 
with spin $\s$ belonging to the irrep 
$\eta$.  Hence, if $\eta'\notin {\cal E}$, $P^{(\eta')}|x\s\rangle=0$ 
and so $P^{(\eta')}|x\ua,x\da\rangle=0$. Therefore, if 
$|\q^{(\eta)}\ket$ is a two-hole singlet eigenstate of the kinetic 
term and $\eta\notin {\cal E}$, then it is also an eigenstate of 
$W_{\mathrm{Hubbard}}$ with vanishing eigenvalue, that means a $W=0$ 
pair.

There are cases when this theorem puts useful restrictions on the possible
ground state symmetries; this happens when the ground state with $N$ 
holes corresponds to filled shells in the $U \rightarrow 0$ limit and 
we wish to study the $N+2$ hole ground state. At least for small $U$ the two added holes 
will form a $W=0$ pair, since this minimizes the ground state energy.
Let ${\cal G}\subset{\cal G}_{0}$ be the symmetry Group of 
$H_{\mathrm{Hubbard}}$ and   
$U_{c}(N)$ the minimum crossover value of $U$, 
that is, the ground state with $N$ holes $|\F_{U}(N)\ket$ has well defined symmetry 
for $0\leq U<U_{c}(N)$. 
Let $|\F_{U=0}(N)\ket$ be non-degenerate 
(e.g. closed shells case) and belong to the one-dimensional irrep 
$\eta_{0}(N)\in{\cal G}_{0}$. 
When we add the two extra holes, the new ground state for $U=0$ is a $W=0$ pair over 
$|\F_{U=0}(N)\ket$, and its symmetry is $\eta_{0}(N)\cdot\eta_{W=0}$, where 
$\eta_{W=0}\in{\cal G}_{0}$ is the symmetry of the added pair. Turning 
on the interaction $\eta_{0}(N)\cdot\eta_{W=0}$ breaks into the direct sum 
of irreps belonging to ${\cal G}$ and among them there must be the 
symmetry of $|\F_{U}(N+2)\ket$ if $0\leq U<U_{c}(N+2)$.  This 
remarkable 
restriction is posed by Group Theory alone: which of the symmetries 
that remain allowed is actually realised in the ground state depends 
on the dynamics. 

The complete characterization of the symmetry of $W=0$ pairs requires 
the knowledge of ${\cal G}_{0}$. A partial use of the 
theorem is possible if one  does not know 
${\cal G}_{0}$ but knows a Subgroup ${\cal G}^{<}_{0}$.
It is then still granted that any pair belonging to an irrep of 
${\cal G}^{<}_{0}$ not
represented in the spectrum has the $W=0$ property.  On the other 
hand, accidental degeneracies occur with a Subgroup of ${\cal 
G}_{0}$, and by mixing degenerate pairs belonging to irreps represented 
in the spectrum one can find $W=0$ pairs also there. 

The theorem tells us that the {\em bigger} is ${\cal G}_{0}$, the 
{\em larger} is the 
number of $W=0$ pairs. Indeed, for a given graph, the number 
of one-body eigenvectors is fixed, while the number and the dimension 
of the ${\cal G}_{0}$ irreps grow with the order of the group. 
Therefore, also the number of irreps not represented in ${\cal E}$ 
grows, and this means more $W=0$ pairs. At the same time, a big 
``non-interacting'' symmetry Group ${\cal G}_{0}$ grants large level 
degeneracies. As we have seen, this fact allows the 
existence of $W=0$ pairs formed by degenerate orbitals. In the next 
Section we show that an anomalously low effective repulsion takes place in 
the interacting system when such levels are on the Fermi surface  
and that in a certain region of the parameters space 
this leads to pairing.

Finally we observe that the $W=0$ theorem predicts $W=0$ pairs of 
$A_{2}$ and/or $B_{2}$ symmetry in the CuO$_{4}$ cluster, as one can 
see by inspection of Figure \ref{1holespectrumcuo4}. This is 
consistent with the above results where the symmetry of the $W=0$ pair 
in Eq.(\ref{ciwfincuo4}) was estimated to be $B_{2}$. Further, when 
$t_{pp}=0$, the non-bonding levels $e$ and $b_{1}$ in Figure \ref{1holespectrumcuo4} 
become degenerate and a new $W=0$ pair of $A_{1}$ symmetry is obtained. 
At a first sight, this fact seems to contradict the $W=0$ theorem, but 
it is not so. As we have observed before, unexpected symmetries arise 
if one use a Subgroup of ${\cal G}_{0}$. For $t_{pp}=0$, the 
extra degeneracy cannot be explained in terms of the $C_{4v}$ Group, whose irreps have 
dimensions 2 at most. It is not hard to realize that for $t_{pp}=0$ 
any permutation of the four Oxygen sites is actually a symmetry and 
therefore the symmetry Group is enlarged to $\P_{4}$ (the group of 
permutations of four objects) which admits irreps of dimension three. 
Then, labelling the one-body levels with the irrep of $\P_{4}$, one 
readly realize that irreps not contained in the spectrum give $W=0$ 
pairs, according to the $W=0$ theorem, and that 
among them there is one containing the $A_{1}$ irrep of $C_{4v}$.}

\subsection{CI and Exact Solutions for Symmetric Clusters}
\label{ciandexsol}

{\small 
In the present Section, we wish to show how $\D<0$ may be induced by 
symmetry through degeneracy.

\subsubsection{$\D<0$ in the CuO$_{4}$}
\label{dnegacuo4}

{\small 
Let us first consider the  CuO$_{4}$  cluster 
with four holes in the non-interacting limit: according to the aufbau 
the first two holes go 
into a bonding level of $A_{1}$ symmetry and the next two into a 
non-bonding level of $E$ symmetry. Group Theory predicts that the 
interactions resolves the sixfold-degenerate ground state $E\otimes E$ 
into $^{3}A_{2}$, $^{1}A_{1}$, $^{1}B_{1}$ and $^{1}B_{2}$. The two- 
body calculations say that $^{1}B_{2}$ and the triplet are 
lowest, since $W=0$ for both. The ground state will turn out to be 
singlet, while  Hund's rule would have predicted  $^{3}A_{2}$;
 we shall see why and how low spin prevails.  

In Table \ref{gsehfcie} we report the 
ground state energies $E_{0}(N)$ of the CuO$_{4}$ cluster with $N$ 
holes, $N=1$ to 4, and the effective repulsion energy 
$\D(4)=E_{0}(4)-2E_{0}(3)+E_{0}(2)$.
\begin{table}[tbp]
    \centering
    \caption{{\footnotesize Ground state energy $E_{0}(N)$ of the CuO$_{4}$ cluster 
    with $N=1$ to 4 holes and effective repulsion 
    $\D(4)=E_{0}(4)-2E_{0}(3)+E_{0}(2)$ between the holes of $E$ 
    symmetry in eV. HF: Hartree-Fock approximation; CI: configuration 
    interaction; Exact: numerical diagonalization of the Hamiltonian. 
    The parameters are those reported in the Table \ref{parameters} 
    except for $U_{pd}$ that is taken to be zero. }}
    \vspace{0.5cm}
   \begin{tabular}{|c|c|c|c|}
      \hline 
               &    HF    &    CI    &    Exact    \\
      \hline 
      $E_{0}(1)$ &     -     &     -     &     -1.72312 \\
      \hline 
      $E_{0}(2)$ &  -1.30523 &     -    &     -1.65603 \\
      \hline 
      $E_{0}(3)$ &  -2.88605  &    -     &  2.21855   \\
      \hline 
      $E_{0}(4)$ &  8.52      &  7.01287 & 6.1006  \\
      \hline
      $\D(4)$    &   1.46    & -0.064   & 0.00747   \\
      \hline 
\end{tabular}
    \label{gsehfcie}
\end{table}
While $W$ gives the direct interaction between the two holes in the 
CI calculation, $\D(4)$ contains indirect interaction as well. The HF 
results correspond to a spontaneously broken symmetry of the 4-hole 
singlet and fail to reproduce the quantum effect we are discussing. 
Accordingly, the effective on-site repulsion is $\D(4)=1.46$ eV, which 
is quite a large repulsive barrier to overcome before pairing can be attained. 
In the CI calculation it is assumed the form (\ref{ciwfincuo4}) for the 
wave function and optimized the energy with respect to the orbitals. 
This does lead to a small negative estimate for $\D(4)$. 
The exact diagonalization with the same parameters yields a positive but tiny  $\D(4)$
with a  $^{1}B_{2}$ ground 
state  (the symmetry of the $W=0$ pair), while the first 
excited state is a $^{3}A_{2}$ triplet lying $D=18$ meV above it. This 
suggests the interpretation that $W=0$ two-hole eigenstates are bare 
quasiparticles and their attractive screened interaction produces the  
pairing\footnote{We note that the experimental evidence on the 
symmetry of the pairs is still conflicting, and different techniques 
tend to give different results. Angular resolved 
photoemission data  favour $d$ pairs with an order parameter of 
$^{1}B_{1}(x^{2}-y^{2})$ 
simmetry. The $W=0$ pair of symmetry $^{1}B_{2}(xy)$  is consistent 
with these findings, since its total-symmetric charge density has 
nodes at 45 degrees from the Cu-O axis. Phases are less significant, 
and we can switch irreps by a gauge transformation. In the full plane the present 
approach also predicts $W=0$ pairs of different symmetries, and the 
symmetry of the most bound pairs depends on the filling. It is fair to 
say that this 
problem is far from being settled. }. 
Although the CI calculation gives a negative $\D(4)$ value, the exact 
calculation yelds a tiny positive one. However, if $t_{pp}$ is 
increased, $\D(4)$ does become negative, which shows that the 
correlation effects may lead to pairing (see Table \ref{deltad}). 
\begin{table}[tbp]
    \centering
    \caption{{\footnotesize Effective interaction $\D(4)$ and energy 
    $D$ of the first excited triplet relative to the singlet ground 
    state as a function of $t_{pp}$ for the CuO$_{4}$ cluster. The 
    off-site $U_{pd}$ parameter is still ignored.}}
    \vspace{0.5cm}
   \begin{tabular}{|c|c|c|c|c|}
      \hline 
        $t_{pp}$ (eV)   &  -0.45    &    -0.25    &   -0.05 & 0    \\
      \hline 
      $\D(4)$ &  -0.0049 &     -0.0209     &   -0.0336 & -0.0364 \\
      \hline 
      $D$ &  0.043 &   0.068  &  0.0915 & 0.097 \\
      \hline 
\end{tabular}
    \label{deltad}
\end{table}
Even if the 
estimation of the singlet-triplet gap depends on the parameters, it 
is consistent with the 45 meV excitation observed in High Resolution 
Electron Energy Loss spectra\cite{demuth} and the 41 meV resonance 
found in Polarized Neutron Scattering experiments\cite{rossat2} on 
superconducting YBa$_{2}$Cu$_{3}$O$_{7}$.

The maximum binding energy occurs at $t_{pp}=0$, when all the 
non-bonding orbitals are degenerate and the configuration mixing is 
thereby enhanced. On the other hand, at positive $t_{pp}$'s, $\D(4)$ 
becomes large and positive, because the $b_{1}$ non-bonding level is 
pushed below the degenerate one, and the mechanism is hampered (at 
$t_{pp}=+0.65$, $\D(4)=0.53$ eV). 

Next, we discuss the dependence of $\D(4)$ on the other 
parameters\cite{cb1}. If 
we decrease $\e_{p}$, $\D(4)$ decreases because this makes the system 
more polarizable. The $\e_{p}$ dependence when all the other parameter 
are kept fixed and $U_{pd}=0$ is almost linear down to $\e_{p}=0$. By 
contrast, the dependence of $\D(4)$ on the Cu-O hopping parameter 
$t_{pd}$ is complex. For small $t_{pd}$, $\D(4)$ is large and 
positive, then falls to zero at $t_{pd}\sim 0.7$ eV. Negative $\D(4)$ 
values are obtained for $t_{pd}\geq 1.4$ eV. Finally let us 
consider the effects of a non-vanishing $U_{pd}$. According to 
Ref.\cite{alascio} positive $U_{pd}$ values do not 
spoil the mechanism, and tend to be synergic with it. Indeed, 
values of $U_{pd}>0.6$ eV give negative $\D(4)$ 
values even for $t_{pp}=-0.65$ eV (in the range 0.2-1.2 eV considered 
in Ref.\cite{cb1} $\D(4)$ is a monotonically decreasing function of 
$U_{pd}$). 

As we have seen, symmetry plays a key role in the analytical calculations; the 
numerical results on the CuO$_{4}$ cluster further demonstrates the 
fact that any lowering of the symmetry is reflected by a corresponding 
increase of $\D(4)$\cite{cb1}. This implies 
that the CI mechanism is working thanks to the $C_{4v}$ 
symmetry, and it is cut off whenever the symmetry is lowered.}

\subsubsection{$\D<0$ in Symmetric Clusters}
\label{dnegasymclu}

{\small 
The total hole concentration which corresponds experimentally to the 
superconducting state is $\r_{h}\sim 0.4$ per atom. In the CuO$_{4}$ 
cluster we need four holes to operate the CI mechanism, because we 
need holes in degenerate states. Therefore $\r_{h}$ is too large by a 
factor of two, and the $W=0$ pair involves nonbonding states, which 
are far from the Fermi level in the bulk.
It is important to realize that these undesirable features are 
peculiar of  the  prototype CuO$_{4}$ cluster, and already disappear
in Cu$_{5}$O$_{4}$, the next larger 
cluster of the same symmetry, see Figure \ref{symclusters}. 
In fact, four holes are still sufficient 
to reach degenerate states, as shown in Figure \ref{cu5o4} ($a$), 
but $\r_{h}\sim 0.44$ is much closer to the experimental value. 

The degenerate states are bonding, in 
contrast with the CuO$_{4}$ case, and the holes now spend a large 
fraction of their time on Cu sites. The effective interaction $\D(4)$ 
for Cu$_{5}$O$_{4}$ and the spin triplet separation $D$ {\it vs.} 
$t_{pp}$ is shown in Figure \ref{cu5o4} ($b$). $\D(4)$ is negative in 
a wide parameters range by a correlation mechanism that does not need 
off-site $U_{pd}$ to operate. To visualize the physical effects of the 
effective attraction, one can compare the ground-state charge 
distribution for the non-interacting and interacting 
holes\cite{cb34}. When $\D(4)<0$, the charge distribution shrinks;
besides, the holes tend to 
accumulate more on the Oxygens where their mutual repulsion is 
mitigated by the CI mechanism. The opposite takes place in the 
``unpaired'' region where the charge of the two least bound holes 
piles up preferentially on the external Cu sites.
The size of the problem with four holes is 1296; it reduces to 81 on projecting to 
the $^{1}B_{2}$ symmetry of the ground state, while the symmetry of 
the first excited state is a $^{3}A_{2}$ triplet. As in the  CuO$_{4}$ 
cluster the maximum binding energy occurs at $t_{pp}=0$\cite{cb2} and 
takes the value $\D(4)=-15.7$ meV; thus the binding energy $|\D(4)|$ is about half 
the CuO$_{4}$ value. For larger clusters the computational effort is 
more demanding (44.100 configurations for the Cu$_{5}$O$_{16}$). For 
$t_{pp}=0$, and $U_{pd}=0$, it was found\cite{cb34} that $\D(4)=-7.5$ meV 
in the Cu$_{5}$O$_{16}$ planar cluster (with $U_{pd}=1$ eV, $\D(4)=-7.6$ 
meV). The calculated trend of $\D(4)$ with the number of atoms $|\L|$ 
is shown in Figure \ref{TrendDelta} ($a$). The decline of $\r_{h}$ with increasing 
number of atoms is slow enough to allow a significant contribution to 
the binding energy for cluster sizes that correspond to the actual 
size ($\x\sim 10-20$ $\stackrel{\circ}{A}$) of the superconducting pair. 
\begin{figure}[H]
\begin{center}
	\epsfig{figure=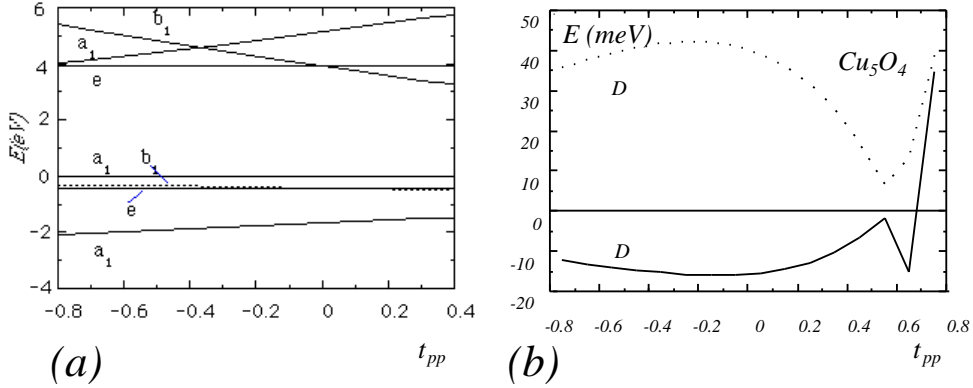,width=13cm}
	\caption{\footnotesize{
	($a$) One-electron energy levels in  
    the Cu$_{5}$O$_{4}$ cluster versus t$_{pp}$. The levels are labelled according
to the irreducible representations of the $C_{4v}$ Group. We find a 
couple of $E$ symmetry (twofold degenerate) levels (one bonding and 
one antibonding), pairs of bonding-antibonding levels of $A_{1}$ and 
$B_{1}$ symmetry and a nonbonding level $a_{1}$. ($b$) 
Dependence of $\D(4)$ and $D$ on $t_{pp}$ for the Cu$_{5}$O$_{4}$ 
cluster. In both ($a$) and ($b$) the parameters (in eV) 
are $\e_{p}-\e_{d}=3.5$, $t_{pd}=1.3$, $U_{p}=6$, 
$U_{d}=5.3$ and $U_{pd}=0$.}}
    \label{cu5o4}
\end{center} 
\end{figure}
As shown in Figure \ref{cu5o4} ($a$),  there are two degenerate 
levels available to test the $W=0$ pairing mechanism; moreover 
 electron pairs and hole pairs are related by 
an approximate charge conjugation symmetry and we can perform the 
test by filling the levels from bottom to top and the other way.
The Cu$_{5}$O$_{4}$ Hamiltonian was  diagonalized\cite{cbs1} and the ground state 
energy was obtained as a function of the number $N$ of holes. 
Eventually, we find electron pairing with a binding energy $\D(N)$ of 
a few tenth of meV in the physical parameter space. 

Let us consider the three-band Hubbard Hamiltonian of 
Eq.(\ref{tbhh}). By a canonical transformation from holes to 
electrons, $c_{s\s}\ra c^{\dag}_{s\s}$, we get for $U_{pd}=0$  
\begin{equation}
H=\sum_{s}(2\e_{s}+U_{s})-\sum_{s\s}(\e_{s}+U_{s})c^{\dag}_{s\s}c_{s\s}-
\sum_{\bra s,s'\ket\s}t_{ss'}c^{\dag}_{s\s}c_{s'\s}+\sum_{s}U_{s}
\hat{n}_{s\ua}\hat{n}_{s\da}.
\end{equation}
Thus, in the electron representation the signs of the site and 
hopping integrals are reversed. With respect to the hole case, the 
sequence of single-electron energy levels of the Cu$_{5}$O$_{4}$ 
cluster is such that bonding and antibonding states are interchanged. 
 In the hole representation the antibonding state of $e$ symmetry contains two holes when
12 holes are present in the cluster. Therefore $N_{h}$=4, 12 are 
magic numbers (to avoid 
confusion we will denote by $N_{h}$ ($N_{e}$) the number of holes  
(electrons)). In the electron representation 
$W=0$ electron pairs are obtained for $N_{e}=4$, 12. Letting 
$E_{0}^{(e)}(N_{e})$ be the ground state energy of the cluster with 
$N_{e}$ electrons, the effective interaction between the two electrons 
in the pair is measured by 
\begin{equation}
\Delta_{e} 
(N_{e})=E_{0}^{(e)}(N_{e})+E_{0}^{(e)}(N_{e}-2)-2E_{0}^{(e)}(N_{e}-1).
\end{equation}
Since the dimensionality of the one-body basis is 18
\begin{equation}
\Delta_{e} (4)=E_{0}^{(e)}(4)+E_{0}^{(e)}(2)-2E_{0}^{(e)}(3)=
E_{0}^{(h)}(14)+E_{0}^{(h)}(16)-2E_{0}^{(h)}(15)=\Delta_{h} (16)
\end{equation}
where $E_{0}^{(h)}(N_{h})$ is the ground state energy of the cluster 
with $N_{h}$ holes. Similarly one finds $\Delta_{e} (12)=\Delta_{h} (8)$. 
\begin{figure}[tbp]
\begin{center}
	\epsfig{figure=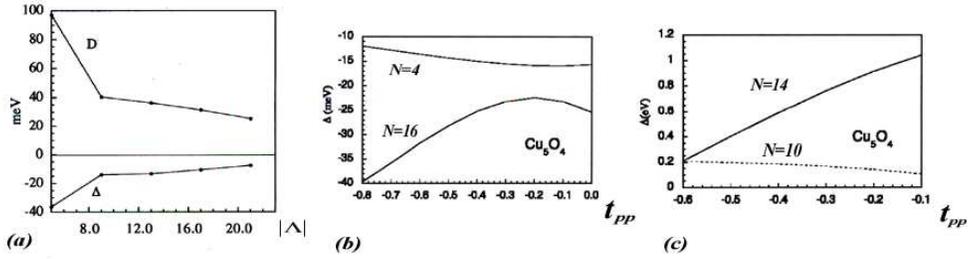,width=13.5cm}
	\caption{\footnotesize{
($a$) $\D(4)$ and the singlet-triplet 
    separation $D$ for allowed clusters containing $|\L|$ atoms. The 
    dots are the values calculated with the parameters listed in the 
    Table \ref{parameters} but $t_{pp}=0$. The curves are a 
    guide to the eye. The clusters are CuO$_{4}$, Cu$_{5}$O$_{4}$, 
    Cu$_{5}$O$_{8}$, Cu$_{5}$O$_{12}$ and Cu$_{5}$O$_{16}$. ($b$) 
    Pairing energy $\D(N)=E_{0}(N)+E_{0}(N-2)+2E_{0}(N-1)$ of the 
    Cu$_{5}$O$_{4}$ cluster with $N=4$ and $N=14$ as a function of 
    $t_{pp}$. ($c$) $\D(N)$ as a function of $t_{pp}$ for $N=10$ and 
    $N=12$.}}
    \label{TrendDelta}
\end{center} 
\end{figure}

We recall that we speak of {\em electron pairs} when two added 
electrons partially occupy a degenerate state and of {\em hole pairs} 
when the same situation is reached by adding two holes; however the 
final situation is exactly the same. One readily verifies that 
in a canonical transformation from
holes to electrons the
two-hole state  becomes a two electron state of the same form.
Therefore the two-body $W=0$ state is invariant under charge 
conjugation, and if holes are paired, electrons are also paired.
In order to avoid switching all the time between the two equivalent pictures, 
below we discuss everything in terms of holes. 

In the symmetric Cu$_{5}$O$_{4}$ cluster we previously found that
the exact diagonalization of the Hamiltonian matrix with four holes in a
wide range of negative $t_{pp}$ values for the $^{1}B_{2}$ singlet gives a
negative $\D(4)$. Therefore one
expects that $\D(N_{h})$ is small or negative for $N_{h}=$ 4, 8, 12 and 
16 and large and positive for any other $N_{h}$.  

The $t_{pp}$ dependence of $\D(N_{h})$ computed using an enhanced Lanczos
diagonalization routine \cite{cbs1} is shown in Figure \ref{TrendDelta} ($b$) 
and ($c$) for $\D(4)$ and $\D(16)$ and for $\D(10)$ and $\D(14)$, respectively. 
In Table \ref{aragd} we summarize the results for $\D(N_{h})$ when 
the model parameters are fixed according to the Table 
\ref{parameters} but $t_{pp}=0$ and for a wide range of $N_{h}$.  One 
sees that  for $N_{h}=$ 4, 8, 12 and 16 $\Delta$  is much smaller in absolute 
value than for other fillings, as predicted. This confirms that $W=0$  
pairs are involved. 
\begin{table}[tbp]
    \centering
    \caption{{\footnotesize Exact diagonalisation results 
    for $\Delta(N_{h})$ (meV), using 
$U_{p}$=6 eV and $U_{d}$=5.3 eV. For $N_{h}$=4, 8 and 16 pairing takes 
place, and at $N_{h}$=12 the repulsion is drastically reduced. 
For $N_{h}$=6 and 14 the $W=0$ pairs 
are {\em not } involved and the normal repulsion develops.}}
    \vspace{0.5cm}
\begin{tabular}{|c|c|c|c|c|c|c|}
\hline
$N_{h}$ & 4 & 6 & 8 & 12 & 14 & 16 \\
\hline
$\Delta$ (meV) & -15.7 & 1469.2 & -10.85 & 43.72 & 1109.2 & -25.47 \\
\hline
\end{tabular}
\label{aragd}
\end{table}

It is apparent that a considerable degree of symmetry under e-h
exchange exists in the cluster and that the pairing interaction is not
restricted to a single value of doping. The electron-hole symmetry is not
exact, because the one-electron energy level spectrum of 
Figure \ref{cu5o4} ($a$) is not 
invariant when up is exchanged with down and left with right 
($t_{pp}\rightarrow -t_{pp}$ and $E \rightarrow -E$); however one
could have predicted the negative $\D$ and its magnitude by the
approximate symmetry. In real systems electron superconductivity has been
first reported\cite{tokura} in the Nd$_{2}$CuO$_{4}$ doped with Ce with a
maximum $T_{c}$=24 K for optimal 
doping concentration $\d$=0.15, similarly to
more common hole superconductors. 
Nd$_{2-\d}$Ce$_{\d}$CuO$_{4}$ is an example
of high-$T_{c}$ cuprate in which charged carriers are electrons. 

To summarize, in this Section we have shown that a correlation effect 
(namely, the CI mechanism) leads to negative $\D$ values by means of the exact 
removal of the direct Coulomb interaction between pairs of degenerate 
hole states. The CI mechanism is a consequence of the symmetry and 
electronic structure of the CuO$_{2}$ plane, and do not depend on the 
size of the cluster. The concept of $W=0$ pairs is useful already at 
a qualitative level to make predictions about the behaviour of 
interacting many-body systems, and this is quite remarkable and 
unusual. Any large deviation from the full symmetry 
restores the normal on-site repulsion The binding energy $\D$ is of the correct order 
of magnitude (10 meV) for the set of parameters which corresponds to 
the best current estimates for the actual system. In the next Section 
we will definitively remove the ambiguity concerning the  
physical meaning of $\D$. We will show that in fully simmetric 
clusters, as the ones studied above, negative $\D$ corresponds to 
pairing.}

\subsection{The Physical Meaning: Effective Interaction}
\label{effecinter}

{\small
In this Section we show that in the Hubbard Model the negative-$\D$ behaviour 
due to the ``$W=0$''  
mechanism has the physical meaning of hole pairing, due to an 
effective attraction between the holes of the  $W=0$ pair.
The fact that pairing occurs can be estabilished by the following argument\cite{cb34}. 
Consider a system $\S$ consisting of a couple of identical clusters 
(say, CuO$_{4}$) that we denote $\S_{a}$ and $\S_{b}$. Let $\S_{a}$ and 
$\S_{b}$ be placed one on top of the other and suppose the two systems are 
linked by a interplane hopping integral $\t<<\D(4)$ as shown in 
Figure \ref{2xCuO4}. 
\begin{figure}[tbp]
\begin{center}
	\epsfig{figure=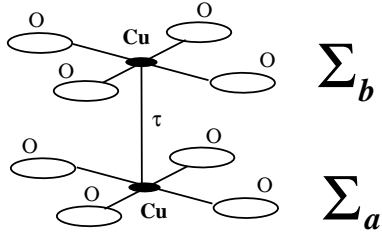,width=5cm}
	\caption{\footnotesize{
Arrangement of the two CuO$_{4}$ planar 
    cluster $\S_{a}$ and $\S_{b}$ described in the text. $\t$ is 
    the hopping integral between the two Cu sites.}}
    \label{2xCuO4}
\end{center} 
\end{figure}
Such an arrangement is chosen arbitrarily for the sake of 
definiteness, but the important thing is that the two clusters are 
linked in such a way that the $C_{4v}$ symmetry is not disturbed and 
the magnitude of $\t$ is such that its influence on the ground state 
energy is negligible compared to the other energies of the system. We 
may then write the full Hamiltonian as $H=H_{\S_{a}}+H_{\S_{b}}+H_{\t}$ and 
regard the weak link term $H_{\t}$ as a perturbation. In zeroth 
order, $H$ can be diagonalized simultaneously with the occupation 
numbers $N_{\S_{a}}$ and $N_{\S_{b}}$ of the two clusters. 
The interesting situation arises 
when the number of holes in $\S$ is $N=N_{\S_{a}}+N_{\S_{b}}=6$. The 
nature of the ground state depends on the parameters. The ``normal'' 
state is obtained if $\D(4)>0$; the unperturbed ground state is 
$|\F_{\mathrm{normal}}\ket=|N_{\S_{a}}=3\ket\otimes|N_{\S_{b}}=3\ket$, 
that is, the product of the individual ground state of two $^{2}E$ 
independent clusters containing three holes each. The
unperturbed ground state energy is  $\approx 2E_{0}(3)$.

On the other hand, if $\D(4)<0$, that means 
$E_{0}(4)+E_{0}(2)<2E_{0}(3)$, the zeroth order ground state is 
\begin{equation}
|\F_{\mathrm{paired}}\ket=\frac{|N_{\S_{a}}=4\ket\otimes|N_{\S_{b}}=2\ket\pm
|N_{\S_{a}}=2\ket\otimes|N_{\S_{b}}=4\ket}{\sqrt{2}}
\end{equation}
and the system prefers to disproportionate. It is clear that $\D(4)$ 
is just the energy needed to break the pair. The $\pm$ sign arises 
because the system has a bonding and an antibonding level; to lowest  
order, the two levels are separated by an energy of the order of 
$\t^{2}/|\D|$. If the system is prepared with a pair of holes in one 
of the planes, it evolves in such a way that the pair will jump back 
and forth from $\S_{a}$ to $\S_{b}$ with that frequency\footnote{We 
remark that interplane oscillations (Josephson plasma mode) had been 
observed in La$_{2-\d}$Sr$_{\d}$CuO$_{4}$, see Ref.\cite{uchida}.}. 
This is clear evidence for pairing; if many systems like $\S$ are 
connected by weak links in a similar way, pairs will 
propagate\footnote{Due 
to the mixing of degenerate states, the ground states of $\S$ 
belonging to positive and negative $\D(4)$ also belong to different 
symmetries. Indeed $|\F_{\mathrm{paired}}\ket$ is nondegenerate, 
since $|N_{\S_{a}}=4\ket$ is $^{1}B_{2}$ while $|N_{\S_{b}}=2\ket$ 
is $^{1}A_{1}$. On the other hand, $^{2}E$ is fourfold degenerate 
and $|\F_{\mathrm{normal}}\ket$ has degeneracy 8 in the $S^{z}=0$ 
subspace.}.

Next, we want to show that in the Hubbard Model the pairing is  due to an 
effective attraction between the holes of the  $W=0$ pair. This point is 
more suble; one could argue that $\D < 0$  depends 
on a comparison of ground state energies of 
systems  with different hole numbers {\it N}, and hence it is not obvious that 
it is a  property of any of them. Moreover, the definition of $\D$ in 
terms of systems with different numbers of particles leads to a 
further difficulty.  When one introduces the vibrational degrees of freedom, the 
state with a sigle hole in the degenerate level can gain energy by a 
Jahn-Teller distortion. The sign of $\D$ 
can even  be reversed. The pairing then looks like an
artifact due to the neglect of the  vibrations. Therefore we want an 
alternative definition of $\D$ which is equivalent to the previous 
one at weak coupling but depends exclusively on the properties of the Hubbard 
Model in the presence of the pair.\footnote{This argument clarifies the 
nature of the pairing in the Hubbard Model, which is due to an 
effective attraction leading to a bound state. Including vibrational 
degrees of freedom in the description of a cluster or a plane, what the 
system would do with more or less holes than it actually has is unimportant if 
we can define the pairing by means of an effective interaction. Much 
more complex 
modeling would of course be necessary to study coupled subsystems like those of
Figure \ref{2xCuO4} in the presence of vibrations. }
We first show how to define the effective interaction in clusters, 
where we may take advantage from the exact diagnalizations.
In all the allowed clusters up to 21 atoms, the lowest one-hole level 
belongs to $A_{1}$ symmetry, 
and the next $e$ level yields the $W=0$ 
pair. In the previous Section we noted that the interactions produce a 
non-degenerate $^{1}B_{2}$ 4-hole ground state with $\D<0$. 
Below, it is shown\cite{cb5} that in the case of dressed $W=0$ singlet
pairs, $\D<0$ arises from an effective pairing interaction which is
attractive; the same interaction is repulsive for triplet pairs. The
argument rests on a comparison of the perturbation series for $\D$ and
for the two-hole amplitude; for the dressed $W=0$ pair the two-hole
amplitude actually involves $\D$ and yields its dynamical
interpretation.} 

\subsubsection{Diagrams for $\D$}
\label{deltadiag}

{\small
The perturbation series for $\D$ in powers of $U$ is obtained by the 
well-known diagrammatic expansion of the ground state energy $E_{0}$ of 
a many body system. This is based on the theorem 
\begin{equation}
E_{0}=E_{0}^{(0)}+i\lim_{t\rightarrow \infty 
(1-i\eta )}\frac{d}{dt}\ln \widetilde{R}(t)  
\label{clustexp}
\end{equation}
where $E_{0}^{(0)}$ is the non interacting ground state energy, $\eta =+0$;
\begin{equation}
\widetilde{R}(t)=\bra\Phi_{0}|U(t)|\Phi_{0}\ket,  
\label{rdit}
\end{equation}
$|\Phi_{0}\ket$ is the non interacting ground state, and $U(t)$ is the time 
evolution operator in the interaction representation. One then expands the 
correlation function $\widetilde{R}(t)$ using the linked cluster 
theorem, 
that simplifies the expansion, but the diagrams that violate the Pauli 
principle and/or the number of particles must be retained. The 
diagrammatic 
rules are readily applied to the $N=2$ case, when $|\Phi_{0}\ket$ is a non 
degenerate single-determinant state. For $N=3$ and $N=4$, the non 
interacting ground state is degenerate, while the derivation of  
Eq.(\ref{clustexp}) 
assumes that the interactions do not modify the symmetry of the ground 
state; therefore we take $|\Phi_{0}\ket$ as the $x$ component of the $^{2}E$ 
irreducible representation for $N=3$ and $^{1}B_{2}$ for $N=4;$ we know from 
direct diagonalization that these symmetries are correct for the whole 
series of clusters. For $N=4$ the $^{1}B_{2}$ component of the non interacting 
ground state reads 
\begin{equation}
|\Phi_{0}\ket=\frac{|e_{x}e_{y}\ket+|e_{y}e_{x}\ket}{\sqrt{2}}  
\label{4hnigs}
\end{equation}
where in terms of orbitals $|e_{x}e_{y}\ket$ is the state  
$|a_{\ua}a_{\da}e_{x\ua}e_{y_{\da}}\ket$. 
Since this cannot be written as a single determinant, some 
care is necessary in the diagrammatic expansion; the 
correlation function is 
\begin{equation}
\widetilde{R}(t)=\bra e_{x}e_{y}|U(t)|e_{x}e_{y}\ket
+\bra e_{y}e_{x}|U(t)|e_{x}e_{y}\ket.  
\label{rtildecorfunc}
\end{equation}
Applying the linked cluster theorem, the first term leads to the
standard diagrams, while the second is anomalous, in the sense that it
vanishes in the non interacting limit. Thus, one finds all the diagrams
contributing to the expansion for $U(t)$ averaged over 
$|a_{\ua}a_{\da}e_{x\ua}e_{y_{\da}}\ket$, 
plus the anomalous or spin-flip diagrams that have
entering $e_{x\ua}$ and $e_{y\da}$ and outgoing $e_{y\ua}$ and 
$e_{x\da}$ lines\cite{cb5}, see Figure \ref{effint} (A). 
\begin{figure}[H]
\begin{center}
	\epsfig{figure=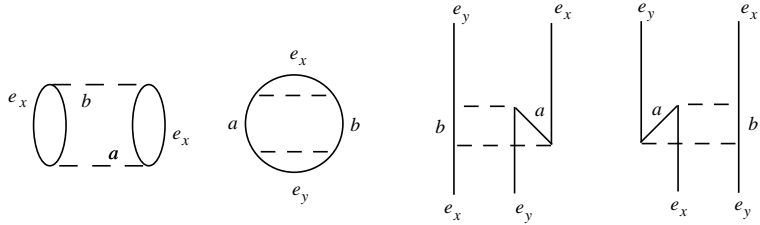,width=10cm}
	\caption{\footnotesize{
(A) Linked diagrams contributing to the second-order 
    expansion of $E_{0}(4)$. Diagram on the left is normal, while the 
    one on the right is anomalous. The $a$, $b$, $e_{x}$ and $e_{y}$ 
    lines are labelled according to the irreps of the $C_{4v}$ 
    Group. (B) Typical spin-flip diagrams for the anomalous propagator 
    $A_{sf}$.}}
    \label{effint}
\end{center} 
\end{figure}
Let $W(k,l,m,n)$ $=\bra k_{\ua}l_{\da}|W_{\mathrm{Hubbard}}|m_{\ua}n_{\da}\ket$; 
the anomalous diagram 
can be obtained from the normal one, which is
proportional to $W(a,b,e_{x},e_{x})^{2}$, simply by letting 
$W(a,b,e_{x},e_{x})\rightarrow
W(a,b,e_{y},e_{y})$ in the upper interaction line. A sign change follows, in
agreement with the standard diagrammatic rules. In summary, one starts the
expansion as if the ground state were the single determinant 
$|a_{\ua}a_{\da}e_{x\ua}e_{y_{\da}}\ket$; the diagrams that involve 
propagating $e_{x}$ and $e_{y}$ 
lines get corrections, which are obtained by exchanging $e_{x}$  
and $e_{y}$ in the upper
interaction line where the $e_{x}$ or $e_{y}$ lines end; 
the loose lines should then be
joined, and the diagram so obtained carries a minus sign because of the
change in the number of loops. It is convenient to denote the hole 
orbitals by $b$ for the $b_{1}$'s and $a$ ($a'$) for the occupied 
(empty) orbitals $a_{1}$. After some algebra, one obtains\cite{cb5} the
second-order approximation to $\D(4)$
\begin{equation}
\D^{(2)}(4)=-2\left[\sum_{b}
\frac{W(a,b,e_{x},e_{x})^{2}}{(\e_{b}-\e_{a})}-
\sum_{a^{\prime }}\frac{W(a,a^{\prime },e_{x},e_{x})^{2}}{(\e
_{a^{\prime }}-\e_{a})}\right] ,  
\label{deltaiiord}
\end{equation}
where $\e_{a}$ is the one-hole energy of the $a$ orbital and so
on; the sums run over all empty states of the appropriate symmetries 
($a_{1}$ and $b_{1}$), while no contributions arise from the empty 
$e$ orbitals since
the relevant $W_{\mathrm{Hubbard}}$ 
matrix elements vanish. Eq.(\ref{deltaiiord}) holds for any symmetric 
cluster up to 21 atoms. 
The sign of $\Delta$ is seen to
depend on the relative weight of the virtual transitions to states of
different point symmetry, and ultimately on the parameters in the
Hamiltonian.}

\subsubsection{Diagrams for the Two-Hole Amplitude}
\label{thadiag}

{\small
Let $G_{2}$ denote the two-hole amplitude for holes of opposite spin in the
degenerate $(e_{x},e_{y})$ orbitals. Singlet and triplet arise from the space
spanned by the non-interacting states $|e_{x}e_{y}\ket$ 
and $|e_{y}e_{x}\ket$, which are connected
by $C_{4v}$ operators. In the Nambu formalism, $G_{2}$ is a matrix 
\begin{equation}
G_{2}(\omega )=\left( 
\begin{array}{cc}
A & A_{sf} \\ 
A_{sf} & A
\end{array}
\right) ;  
\label{nambu}
\end{equation}
here $A$ is a normal propagator, while $A_{sf}$ involves a spin-flip and
vanishes in the non-interacting limit. When interactions are included, the
first order contribution to the scattering amplitude vanishes because the
product of $e_{x}$ and $e_{y}$ orbitals vanishes on all sites. 
Therefore, the spin-flip
process, $A_{sf}$, shown in Figure \ref{effint} (B), 
is the lowest-order $e_{x}-e_{y}$ scattering,
and produces the effective interaction to second-order. The system 
makes virtual transition to 4-body (3 holes and 1 electron) states. 

Evaluating $A_{sf}$ according to the standard rules, one finds\cite{cb5} that near 
$\omega =2\e_{e}$, 
$A_{sf}\approx \frac{i\Delta ^{(2)}(4)}{(\omega -2\e_{e})^{2}}$
+less singular terms; moreover, to second order, $A=\frac{i}{\omega
-2\e_{e}}$. We know from symmetry that $G_{2}(\omega)$ has singlet and
triplet eigenvectors $\frac{1}{\sqrt{2}}(1,\pm 1)$; for the singlet and
triplet we get
\begin{equation}
A+A_{sf}=\frac{i}{\omega -\eta_{s}}, \;\;\;\;\; 
A-A_{sf}=\frac{i}{\omega -\eta_{t}},
\label{singtripprop}
\end{equation}
where $\eta_{s,t}$ are the new eigenvalues. Therefore, for the singlet the
expansion is
\begin{equation}
\frac{1}{\omega -\eta_{s}}\approx \frac{1}{\omega -2\e_{e}}+
\frac{\D^{(2)}(4)}{(\omega -2\e_{e})^{2}}\approx \frac{1}{\omega
-2\e_{e}-\D^{(2)}(4)},  
\label{singexp}
\end{equation}
that is $\eta _{s}=2\e_{e}+\D^{(2)}(4)$; the triplet receives
the opposite correction, and the $^{1}B_{2}-^{3}A_{2}$ separation is 
$D=2|\D(4)|$. Negative $\D(4)$ means that the spin-flip interaction is
attractive for $^{1}B_{2}$ which is pushed down by $|\D(4)|$ 
and becomes the ground state, while the triplet is pushed up by the same
amount. In this way, $\D(4)$ can be redefined without any reference 
to the ground state of clusters with a different numbers of holes, and 
we are free from the objections based on the Jahn-Teller distorsion of 
odd-$N$ clusters. 

Since $U_{p}$ and $U_{d}$ are not small compared to the Cu-O hopping term 
$t_{pd}$, 
the second-order is generally a poor approximation; interestingly, $W=0$
pairs are an exception, because the large interactions are {\it dynamically} 
small. The first-order term vanishes, and the second order is of the
order of tens of meV. Comparison with exact numerical diagonalization
results shows that the second-order approximation for $D$ and $\D(4)$ is
already rewarding. In Figure \ref{TrendDelta} ($a$) 
it is shown how $D$ and $\D(4)$ scale with the
cluster size; they are indeed very closely proportional, although their
ratio is somewhat larger than 2. Thus, the comparison with $D$ is 
useful to show that $\D(4)$ really measures a property of the 4-hole 
state, and that the two $E$ symmetry holes actually experience 
attraction. 

\subsection{Superconducting Flux Quantization}
\label{fluxqinsymclusters}

{\small 
We test the paired state by exposing the system to a vector potential 
${\mathbf A}({\mathbf r})$ according to the Peierls prescription,
\begin{equation}
t_{ss'}\rightarrow t_{ss'}\exp [\frac{2\pi i}{\phi _{0}}\int
\limits_{s}^{s'}{\bf A}\cdot d{\bf r}],  
\label{peierlspre}
\end{equation}
where $\phi _{0}=2\p/e$ is the flux quantum, and looking for
superconducting correlations. In a macroscopic experiment, one makes a
sample with a large hole, and inserts a magnetic field; the ground state
energy $E_{0}(\phi )$ is trivially a periodic function of 
$\phi/\phi_{0}$, where $\phi$ is the flux in the tube, and $\phi=0$ 
is an extremum\footnote{A flux $\phi=\phi_{0}$ can be gauged away, 
and any physical property, for example the ground 
state energy, is a periodic functions of $\phi$ with period 
${\phi_{0}}$.}. 
Superconductors quantize the flux by allowing integer and half-integer
multiples of $\phi_{0}$, because the quasiparticles that screen the 
vector potential carry charge $2e$. A flux $\phi_{0}/2$
corresponds to a superconducting ground state with pairs having different
symmetry than those prevailing at $\phi=0$, and the system is stable;
arbitrary $\phi$ values are not allowed because they cost large amounts of
(free) energy.

Canright and Girvin have shown\cite{girvin} that cluster calculations can be
used to obtain qualitative insight on the occurrence of superconductivity,
by looking for a tendency to flux quantization. The 
signature\cite{byers} is
present when $\phi=0$ is a minimum of $E_{0}(\phi)$ and the only other minimum
of comparable depth occurs at $\phi/\phi_{0}=\frac{1}{2}$; the
barriers separating the minima are small in a small system but one expects
them to increase with size, leaving the flux quantized in units of 
$\phi_{0}/2$. Canright and Girvin\cite{girvin} used a square lattice of
rectangular geometry and periodic boundary conditions along one of the axes;
to demonstrate the effect, they assumed an {\it attractive} on-site
interaction and observed superconducting flux quantization for even hole
numbers and strong enough attraction. In the present problem, with a repulsive Hubbard 
Model, the mechanism of attraction is driven by the $C_{4v}$ 
symmetry, and  cannot operate with  such an unsymmetric geometry.  
The flux must be inserted in such a way that the system is not distorted. 
We must insert the flux tube 
inside allowed clusters, but since no closed path encircling the central 
Cu is available for the holes, there is no response to a central flux tube.
\begin{figure}[tbp]
\begin{center}
	\epsfig{figure=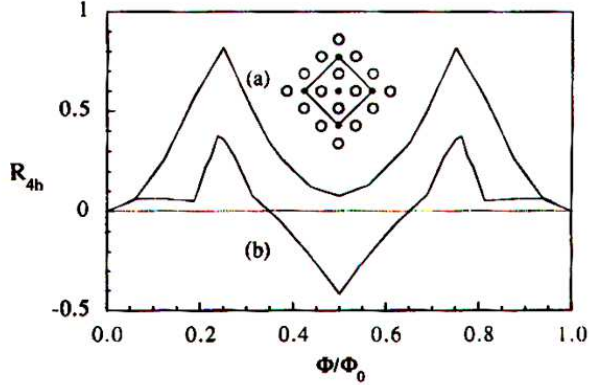,width=8.5cm}
	\caption{\footnotesize{
Four-hole response function 
    $R$ versus the normalized flux, for (a) Cu$_{5}$O$_{4}$ 
    and (b) Cu$_{5}$O$_{16}$. The parameter values are the same of 
    Table \ref{parameters}. The inset shows the geometry of 
    Cu$_{5}$O$_{16}$ with the closed path.}}
    \label{fluxq1}
\end{center} 
\end{figure}
Cini and Balzarotti\cite{cb5} used an {\it external} 
device providing a closed path around
the flux tube. The device, however, cannot be chosen arbitrarily, because
the cluster must remain allowed and the singlet pair must remain one with 
$W=0$. To fulfill this condition, Cini and Balzarotti 
introduce an infinitesimal hopping $t_{dd}$ between the external 
Cu's (see the inset of Figure \ref{fluxq1}) and study the linear
response to this perturbation\footnote{Due to the smallness of the 
system under study, the hopping parameter $t_{dd}$ that contribute a 
closed path around the central Cu must be kept very small to prevent 
the magnetic perturbation from growing large and destroying the 
structure of the ground state multiplet. Numerically, the computations 
were performed with $t_{dd}=\pm 0.01$ eV.}. This geometry is a compromise, 
because the magnetic field penetrates our small cluster; however, it lends 
itself to an extension to the full plane, such that only the 4 central 
plaquettes feel a magnetic field, and the rest of the plane only 
experiences a vector potential (see Section \ref{chapter4}). 
The relevant response function $R$ is 
\begin{equation}
R\left( \phi \right) =\frac{E_{0}(\phi )-E_{0}(0)}{t_{dd}}  
\label{respmagnf}
\end{equation}
and depends on the test flux $\phi$ (in units of $\phi_{0}$). The results
for Cu$_{5}$O$_{4}$ and Cu$_{5}$O$_{16}$ with 4 holes are shown in 
Fig.\ref{fluxq1}; the
same trend is obtained for all clusters. The minimum at $\phi_{0}/2$ 
is the microscopic precursor of the superconducting flux quantization; by
looking at $\Delta(4)$ and analyzing the numerical ground state wave functions
at $\phi_{0}/2$ one finds that it corresponds to pairs of $^{1}A_{1}$
symmetry which are bound, since the minimum at half flux quantum
corresponds again to a $\D(4)<0$ situation. Thus 
the point symmetry of the wavefunction changes from $^{1}B_{2}$($x^{2}-y^{2}$)
at $\phi=0$ to $^{1}A_{1}$($x^{2}+y^{2}$) at $\phi=\phi_{0}$/2. 

This finding is a much clearer signature of
superconducting flux quantisation than the generally accepted presence 
of the two minima, because it implies that the superconductor remains 
a superconductor after swallowing up the half flux quantum. Therefore, 
in order to determine the flux dependence 
of the effective interaction, one can compute  $\Delta(N, \phi)$. It 
was found\cite{epj2000} that {\em when  $\Delta(N, 0)<0$, then
$\Delta(N,\phi_{0}/2)$ is also negative.}

Quite different patterns in which the flux quantization is absent are obtained for
two and three holes\cite{cb5}. Also, it is clear that it is 
$W_{\mathrm{Hubbard}}$ that forces
the paired holes to screen the vector potential like a charge $2e$; actually,
the non interacting case is quite different\cite{cb5}. If the input data 
are modified in such a way that $\Delta$ becomes positive by severe
distortion of the symmetry, the characteristic central minimum is lost
altogether.

By looking at Fig.\ref{fluxq1} we realize that with increasing 
the flux $\phi$, the $^{1}B_{2}$ component of the ground state is quickly 
destabilized by the vector potential, while the $^{1}A_{1}$ component 
decreases in energy and eventually becomes energetically favored in 
presence of the trapped flux. A similar change of symmetry occurs in 
conventional superconductors where the Cooper wavefunction has 
$s$-wave symmetry in absence of magnetic flux. 

These findings are required by general symmetry principles. In the
absence of $t_{dd}$, the full invariance Group of the cluster is $\P_{4}$ 
and, as we have seen above, the interacting ground
state is degenerate, with $^{1}A_{1}$ and $^{1}B_{2}$ components. A
nonzero $t_{dd}$ at $\phi=0$ reduces the symmetry to the $C_{4v}$ Subgroup; 
it turned out numerically that  with a positive $t_{dd}$ ($t_{dd}=0.01$ eV in our 
calculations) the expectation value of 
the magnetic perturbation is negative on $^{1}B_{2}$ and positive on 
$^{1}A_{1}$; therefore  the ground state is $^{1}B_{2}$ at 
$t_{dd}>0$ but changes symmetry if the sign of $t_{dd}$ is reversed. 
Upon switching the vector
potential ${\bf A}$, the Cu-Cu hopping is complex and chiral, so the symmetry is 
lowered again from $C_{4v}$ to its  Subgroup ${\mathbf Z}_{4}$,
which contains only the rotations. Since ${\mathbf Z}_{4}$  is Abelian, there are no
degeneracies for a generic $\phi$, so  there are no $W=0$ pairs and 
therefore 
repulsion prevails. With increasing the flux  from 0, the ground state
energy increases to a maximum. Then it  decreases because, at $\phi 
=\phi_{0}/2$, the Cu-Cu hopping of Eq.(\ref{peierlspre}) becomes 
$-t_{dd}$, which is real; then
the full $C_{4v}$ symmetry is restored, resurrecting the $W=0$ pairs. The 
recovery of $C_{4v}$ at $\phi =\phi_{0}/2$ enables us to 
assign the eigenvectors to the irreps, as noted above. 
The change of symmetry of the pair is 
also readily 
understood: the perturbation caused by $t_{dd}>0$ at $\phi=0$ becomes the 
opposite at half fluxon, so the $^{1}A_{1}$ state is lowest now. 
The signature of superconducting pairing is not only the existence of a well defined 
second minimum at half flux quantum, but also the fact that it corresponds 
to a  pairing ($\Delta <0$ ) situation, like at $\phi=0$. It is gratifying that the
present model allows for $W=0$ pairs of both symmetries, because this is the
only way the superconducting flux quantization works.

Finally, it is worth to note that the Hubbard large $U$ 
system is stabilized by the intrinsic charge fluctuation. 
Since the mechanism is robust (first order) it is conceivable 
that these low lying pairing states can survive in the vortex core of 
the underdoped phase\cite{fischer}.}

\section{Pairing in the Hubbard Model: General Theory}
\label{chapter4}

{\small 
The CuO$_{2}$ plane and the clusters that possess the same $C_{4v}$ symmetry around a Cu 
ion have 2-hole eigenstates of the kinetic 
energy with vanishing on-site repulsion ($W=0$ pairs). 
Cluster calculations by exact diagonalisation show 
that these are the quasiparticles that lead 
to a paired ground state, and have 
superconducting flux-quantisation properties. 
In this Section, we extend
the theory to the full plane, and show that the 
$W=0$ quasiparticles are again the natural
explanation of superconducting flux quantisation. 
Moreover, by a novel canonical tranformation approach, 
we calculate the effective interaction 
$W_{\mathrm{eff}}$ between two holes added to the
ground state of the repulsive Hubbard Model. The method  
is a particularly efficient way to
perform the configuration interaction calculation. It is based on the symmetry, and 
dramatically displays the mechanism of pairing in the CuO$_{2}$ 
plane. 
We show that the full configuration interaction calculation can 
be performed recursively. At each step, one decouples a class of 
virtual excitations while renormalising the matrix elements of the 
kinetic term $K$ 
and of the Hubbard repulsion $W_{\mathrm{Hubbard}}$. 
At the end, one obtains an exact, analytical canonical transformation 
producing an effective Hamiltonian for the dressed pair. 
In order to get actual numbers, however, we 
had to neglect the renormalizations in the final formula; 
this  approximation is fully justified at weak 
coupling. Thus, a closed-form analytic
expression including the effects of all virtual transitions to 4-body intermediate 
states (exchange of an electron-hole pair) is derived. 
The effective interaction is the result of a partial cancellation 
of positive and negative contributions, so it is not necessarily 
attractive in all cases; the general signature of $W=0$ pairs is that 
the absolute value of the interaction is much smaller than in the 
other cases. Since $W=0$ pairs  emerge from  symmetry alone, they remain 
$W=0$ for any coupling strength and are an adequate starting point 
for a realistic theory. This is the reason why  weak coupling 
expansions often provide good approximations at intermediate 
coupling, as observed by several authors\cite{metzner2}\cite{fridman}\cite{galan}.  

The main scheme is ready to include  other interactions which are not considered 
in the Hubbard Model but may be important. We find that the effective interaction 
is attractive and leads to a Cooper-like instability of the Fermi
liquid, while it is repulsive for triplet pairs. From $W_{\mathrm{eff}}$, 
an integral equation for the pair eigenfunction can be derived; the binding energy 
$|\Delta|$ of the pairs is in the range of tens of meV, which is 
not comparable to any of the $U$ and $t$ input parameters. The reason 
is that the interaction, which vanishes identically for the {\em 
bare} $W=0$ pairs, remains {\em dynamically} small for the dressed 
quasiparticles. This suggests that a {\em weak coupling} theory may be 
useful to study the pairing force, despite the fact that $U$ is not
small compared to $t$.

\subsection{Band Structure}
\label{bands}

{\small
Let us consider the three-band Hubbard Hamiltonian. The kinetic term 
is 
\begin{equation}
K=\sum_{\bra s,s'\ket}t_{ss'}c^{\dag}_{s\s}c_{s'\s}+
\sum_{s}\e_{s}c^{\dag}_{s\s}c_{s\s}
\end{equation}
where $t_{ss'}=t_{pd}$ if $s$ and $s'$ represent a Copper site and a 
nearest neighbor Oxygen site, or {\it viceversa}, and zero otherwise, while 
$\e_{s}=\e_{p}$ if $s$ is an Oxygen site and zero otherwise. It is 
convenient to identify each unit cell by its cartesian coordinates 
${\mathbf r}=(i_{x},i_{y}),\;i_{x},i_{y}=1,\ldots,N_{\L}$ 
($N_{\L}^{2}=|\L|/3$ 
being the total number of unit cell) in such 
a way that ${\mathbf r}$, ${\mathbf r}+{\mathbf 
a}$ and ${\mathbf r}+{\mathbf b}$ are the positions of the Cu and of 
the two Oxygens, respectively. The one-body eigenstates of $K$ are 
\begin{equation}
K c^{(\n)^{\dag}}_{{\mathbf k}\s}|0\ket=\e^{(\n)}_{{\mathbf k}}
c^{(\n)^{\dag}}_{{\mathbf k}\s}|0\ket,\;\;\;\;{\mathbf k}=(k_{x},k_{y})=
\frac{2\p}{N_{\L}}(i_{x},i_{y}),\;\;\;\;i_{x},i_{y}=1,\ldots,N_{\L},
\end{equation}
where $\nu$ is a band index and periodic boundary conditions are 
chosen. According to Bloch's theorem they can be 
written as 
\begin{equation}
c^{(\n)^{\dag}}_{{\mathbf k}\s}=\sum_{s}c^{\dag}_{s\s}\q^{(\n)}({\mathbf 
k},s),\;\;\;\;
\q^{(\n)}({\mathbf k},s)=e^{i{\mathbf k}\cdot {\mathbf r}}
\vf^{(\n)}({\mathbf k},s)
\label{cdiag}
\end{equation}
where $\vf^{(\n)}$ has the lattice periodicity and ${\mathbf r}$ is 
the position of the cell in which the site $s$ lies. The non-bonding band is 
characterized
by $\vf^{(nb)}({\mathbf k},0)=0$; one then finds
\begin{equation}
\vf^{(nb)}({\mathbf k},{\mathbf a})=\frac{1}{N_{\L}}
\frac{\cos\left[\frac{k_{x}}{2}\right]}
{\sqrt{\left(\cos^{2}\left[\frac{k_{x}}{2}\right] 
+\cos^{2}\left[\frac{k_{y}}{2}\right]\right)}},
\;\;\;\;\;\;
\vf^{(nb)}({\mathbf k},{\mathbf b})=\frac{1}{N_{\L}}
\frac{\cos\left[\frac{k_{y}}{2}\right]}
{\sqrt{\left(\cos^{2}\left[\frac{k_{x}}{2}\right] 
+\cos^{2}\left[\frac{k_{y}}{2}\right]\right)}}.
\end{equation}
For the other bands, one obtains the eigenvalue equation
\begin{equation}
\e^{\left( \pm \right) }_{{\mathbf k}} =\frac{
\e_{p}\pm \t}{2},
\;\;\;\;\;\;\;\;\;\;\;\;\;
\t=\sqrt{\e_{p}^{2}+16t_{pd}^{2}\left[\cos^{2}\left[ 
\frac{k_{x}}{2}\right]+\cos^{2}\left[\frac{k_{y}}{2}\right] 
\right] }.
\label{scorc}
\end{equation}
(+ for the antibonding and - for the bonding band) 
and the Cu amplitudes 
\begin{equation}
\vf^{\left( \pm \right) }\left( {\mathbf k},0\right) =\frac{1}{N_{\L}}
\sqrt{\frac{\e
^{\left( \pm \right) }-\e_{p}}{\left( 2\e
^{\left(
\pm \right) }-\e_{p}\right) }},
\end{equation}
while the O amplitudes are obtained from
\begin{equation}
\left( \e_{p}-\e^{(\pm)}_{{\mathbf k}}
\right) \vf^{(\pm)}\left( {\mathbf k},{\mathbf a}\right)
 +2t_{pd}\cos \left[ \frac{
k_{y}}{2}\right] \vf^{(\pm)}\left( {\mathbf k},0\right) =0,
\end{equation}
and
\begin{equation}
\left( \e_{p}-\e^{(\pm)}_{{\mathbf k}}
\right) \vf^{(\pm)}\left({\mathbf k},{\mathbf b}\right)
+2t_{pd}\cos
\left[ \frac{ k_{x}}{2}\right] \vf^{(\pm)}\left(
{\mathbf k},0\right) =0.      
\end{equation}

To write $W_{\mathrm{Hubbard}}$ in terms of the $c^{(\n)^{\dag}}_{{\mathbf k}\s}$ 
we must invert Eq.(\ref{cdiag}): $c^{\dag}_{s\s}=
\sum_{{\mathbf k}\n}c^{(\n)^{\dag}}_{{\mathbf k}\s}
\q^{(\n)^{\ast}}({\mathbf k},s)$. 
Thus,
\begin{equation}
W_{\mathrm{Hubbard}}=\sum_{{\mathbf k}_{1}..{\mathbf k}_{4}}\sum_{\nu_{1}..\nu_{4}}
U^{\left(\nu_{1}..\nu_{4}\right)}\left({\mathbf k}_{1},{\mathbf k}_{3},
{\mathbf k}_{2},{\mathbf k}_{4} \right)
c^{(\nu_{1})^{\dag}}_{{\mathbf k}_{1}\ua}c^{(\nu_{2})}_{{\mathbf k}_{2}\ua}
c^{(\nu_{3})^{\dag}}_{{\mathbf k}_{3}\da}c^{(\nu_{4})}_{{\mathbf 
k}_{4}\da}. 
\label{wk}
\end{equation}
Using a shorthand notation, where the band indices are understood,
\begin{equation}
U\left({\mathbf k}_{1},{\mathbf k}_{3},
{\mathbf k}_{2},{\mathbf k}_{4} \right)=\sum_{{\mathbf r}}
e^{-i({\mathbf k}_{1}+{\mathbf k}_{3}-{\mathbf k}_{2}-{\mathbf k}_{4})
\cdot {\mathbf r}}\sum_{s\in\mathrm{cell}}U_{s}
\vf^{*}\left({\mathbf k}_{1},s \right)\vf^{*}\left({\mathbf k}_{3},s \right)
\vf\left( {\mathbf k}_{2},s
\right)\vf\left( {\mathbf k}_{4},s \right), 
\label{uk}
\end{equation}
where $\sum_{s\in\mathrm{cell}}$ involves a sum over a single cell. 
The umklapp
processes involving the basis vectors of the reciprocal lattice   
produce a minus 
sign on one of the Oxygens. The matrix elements (\ref{uk}) are 
proportional to $N_{\L}^{-2}$, like any {\em bona fide} two-body 
interaction.

The one-band Hubbard Model is even more elementary; we postpone the 
details to the next Section.}

\subsection{$W=0$ pairs and Superconducting Flux Quantization}
\label{w=0andflux}

{\small
In this Section we demonstrate that $W=0$ pairs with vanishing 
momentum and formed by holes in degenerate states exist whatever is 
their energy. Thus, in the thermodinamic limit, there are no 
constraints on the occupation number; any even number is ``magic''. 
Instead of using the $W=0$ theorem and the full symmetry, that would in principle yield all 
the $W=0$ pairs, we follow the simpler route of  projecting a single
determinantal state on the irreps of  $C_{4v}$ \cite{epj1999}\cite{ssc1999}. 
This partial use of the symmetry still gives enough solutions to demonstrate 
pairing.

Let us first consider the Hubbard Model with 
three bands. Omitting  band indices again, we shall mean 
\begin{equation}
|d[{\mathbf k}]\ket=c^{\dagger}_{{\mathbf k},\ua}
c^{\dagger}_{-{\mathbf k},\da}|0\ket 
\label{detercoop}
\end{equation}
to be a two-hole determinantal state derived from the ${\mathbf k}$ 
eigenfunctions. 
Since $\vf\left(-{\mathbf k},s\right)=\vf\left({\mathbf k},s\right)$, which is required 
by time-reversal symmetry, the combination $|d[{\mathbf k}]\ket+|d[-{\mathbf k}]\ket$ 
is singlet and $|d[{\mathbf k}]\ket-|d[-{\mathbf k}]\ket$ is triplet. 

We introduce the determinants
$|d[R{\mathbf k}]\ket\equiv|d[{\mathbf k}_{R}]\ket,\;R\in C_{4v}$, and the projected states 
\begin{equation}
|\Phi_{\eta}\left[ {\mathbf k}\right] \ket=\frac{1}{\sqrt{8}}{\sum_{R\in 
C_{4v}}}\chi
^{\left( \eta \right) }\left( R\right)|d[{\mathbf k}_{R}]\ket  
\label{detrproj}
\end{equation}
where $\chi^{\left( \eta \right) }(R)$ is the character of the 
operation $R$ in the irreducible representation $\eta$. In the 
non-degenerate irreps, the operations that produce opposite 
${\mathbf k}_{R}$ have the same character,
and the corresponding projections lead to singlets. Let 
$R_{i},i=1,..8$
denote the operations of $C_{4v}$ and ${\mathbf k},{\mathbf k}^{\prime}$ 
any two vectors in the Brillouin Zone (BZ).
Consider any two-body operator $\hat{O}$, which is symmetric
($R_{i}^{\dagger}\hat{O}R_{i}=\hat{O}$), and the matrix with elements 
$O_{i,j}=\bra d[{\mathbf k}]|R_{i}^{\dagger}\hat{O}R_{j}|d[{\mathbf 
k}']\ket$, where ${\mathbf k}$ and ${\mathbf k}^{\prime}$ 
may be taken to be in the same or in different bands. This matrix 
can be
diagonalized by Group Theory alone. Indeed, for each irrep 
$\eta$, consider the normalized vector with components 
$\frac{1}{\sqrt{8}}\chi ^{\left( \eta \right) }\left( R_{i}\right)$; 
this is an eigenvector of the operator matrix 
$R_{i}^{\dagger}\hat{O}R_{j}$, as we
can check by noting that
$\frac{1}{8}\sum_{i,j}\chi^{\left(\eta\right)}\left(R_{i}\right)
R_{i}^{\dag}\hat{O}R_{j}\chi^{\left(\eta'\right)}\left(R_{j}\right)
=\hat{O}P^{(\eta)^{2}}\delta\left(\eta,\eta'\right)$; 
hence, $R_{i}^{\dagger}\hat{O}R_{j}$ is diagonal on the basis of the 
symmetry
projected states. The square of the projection operator is
$P^{(\eta)^{2}}=\sqrt{8}P^{(\eta)}$. 
Now, taking the matrix element between the ${\mathbf k}$ and ${\mathbf k}^{\prime}$  
determinants, we get the eigenvalues in terms of determinantal matrix elements:
\begin{equation}
O\left( \eta ,{\mathbf k},{\mathbf k}^{\prime }\right) =
{\sum_{R}}\chi ^{\left( \eta 
\right)}\left( R\right) O_{R}\left( {\mathbf k},{\mathbf k}^{\prime }\right)  
\label{os1}
\end{equation}
where $O_{R}\left( {\mathbf k},{\mathbf k}^{\prime }\right) =
\langle d[{\mathbf k}|\hat{O} |d[{\mathbf k}^{\prime
}_{R}] \rangle$. If
$\hat{O}$ is identified with $W_{\mathrm{Hubbard}}$, the determinantal matrix element 
does not depend on the sign of the components of
${\mathbf k}^{\prime}$, because the
$\vf$'s depend on ${\mathbf k}$ only through a cosine. The only thing 
that matters is that $C^{(+)}_{4},\;C_{4}^{(-)},\sigma_{+}$ and
$\sigma_{-}$ exchange $k_{x}^{\prime}$ and $k_{y}^{\prime}$, 
while the other operations leave them in place. Consequently, 
$(W_{\mathrm{Hubbard}})_{{\mathbf 1}}$ = $(W_{\mathrm{Hubbard}})_{C_{2}}$ = $
(W_{\mathrm{Hubbard}})_{\sigma _{x}}$ = $(W_{\mathrm{Hubbard}})_{\sigma _{y}}$ and
$(W_{\mathrm{Hubbard}})_{C^{(+)}_{4}}$ = $(W_{\mathrm{Hubbard}})_{C_{4}^{(-)}}$ = $
(W_{\mathrm{Hubbard}})_{\sigma _{+}}$ = $(W_{\mathrm{Hubbard}})_{\sigma_{-}}$. 
Therefore 
\begin{equation}
W_{\mathrm{Hubbard}}\left( ^{1}A_{2}\right) =
W_{\mathrm{Hubbard}}\left( ^{1}B_{2}\right) =0 .
\end{equation}
These are $W=0$ pairs, like those studied previously\cite{cb5}, but 
with an important change, since in the cluster calculations the symmetries of 
$W=0$ pairs were found\cite{cbs1} to be
$^{1}B_{2}$ and $^{1}A_{1}$. The reason for this change is a twofold 
size effect. On one hand, $^{1}A_{1}$ pairs have the $W=0$ property only in the 
small clusters,
having the topology of a cross, and belonging to the $\P_{4}$ Group, 
but do not
generalize as such to the full plane, when the symmetry is lowered to 
$C_{4v}$; on
the other hand, the small clusters admit no solutions of
$^{1}A_{2}$ symmetry at all if only degenerate states are 
used.
One necessary condition for pairing in clusters is that the least bound 
holes form such a pair, and this dictates conditions on the occupation number. 
In the full plane, however, $W=0$ pairs exist at the Fermi level for any filling. 

Let us now explore what happen if a magnetic field is turned on\cite{epj2000}.  
Consider the pattern of Figure \ref{magnfinplane} (A).
\begin{figure}[tbp]
\begin{center}
	\epsfig{figure=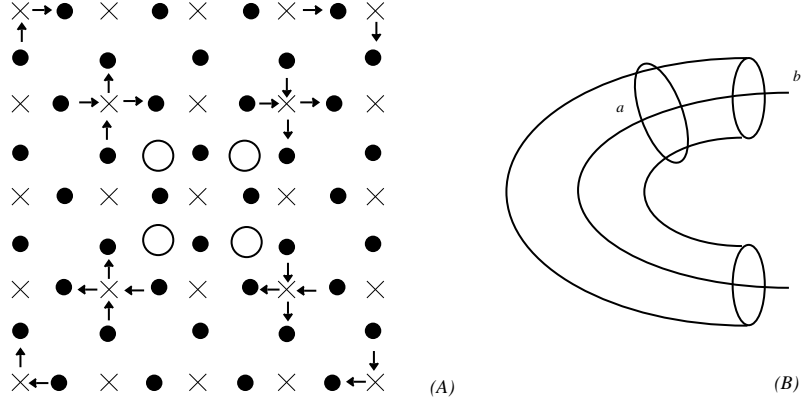,width=10.5cm}
	\caption{\footnotesize{
(A) Pattern of the vector potential ${\mathbf A}$ 
    due to 4 flux tubes (empty dots) carrying flux $\phi$. X stands for 
    Cu. The line integral of ${\mathbf A}$ along each bond parallel to the arrow is 
    $\f/2$. (B) The geometry of the torus and of the magnetic field used to discuss 
    the flux quantisation properties of the $W=0$ pairs.}}
    \label{magnfinplane}
\end{center} 
\end{figure}
Here, the Copper sites are marked by X and the 
Oxygen sites by O; the black dots stand for tubes carrying flux 
$\phi$ each, symmetrically disposed around the central Cu. Varying 
$\phi$ by an integer multiple of $\phi_{0}$ corresponds to a gauge 
transformation leaving all the physical properties invariant. The arrows 
help to visualise a convenient choice of the vector potential at general $\phi$. 
Namely, running along an oriented bond in the sense of the arrow, 
\begin{equation}
\int_{\rightarrow} {\bf A}\cdot d{\bf r}=\frac{\phi}{2};
\label{arrows}
\end{equation}
along the other Cu-O bonds, not marked in the 
figure, $\int {\bf A}\cdot d{\bf r}=0$. One sees that in this way the flux through any 
closed path corresponds to the number of tubes surrounded by the path. 
The reflection operations of $C_{4v}$ are equivalent to 
$\phi\rightarrow -\phi$, reverse the directions of the 
arrows and for a generic $\phi$ the symmetry Group reduces to 
${\mathbf Z}_{4}$. However, at $\phi=\phi_{0}/2$ the reversal of the 
magnetic field in the tubes corresponds to a jump by $\phi_{0}$, and 
this is equivalent to a gauge transformation: this implies that the symmetry 
Group gets bigger, the new symmetry operations being reflections 
supplemented by a gauge transformation. Indeed, it follows from 
Eq.(\ref{arrows}) that the hopping parameter becomes $it_{pd}$ along the 
arrows, while it remains equal to $t_{pd}$ along the unmarked bonds of 
Figure \ref{magnfinplane} (A). Any reflection operation simply 
changes the signs of all the hoppings along the marked bonds. Now 
consider the unitary transformation ${\cal U}$ which changes the signs of all
the Cu orbitals along both diagonal, except the central Cu. Since ${\cal U}$ 
also has the effect of reversing all the arrows, $\sigma\cdot{\cal U}$ is a 
symmetry, for all reflections $\sigma$ in $C_{4v}$.  Moreover, since 
the product of two reflections is a rotation, the Group $\tilde{C}_{4v}$
including the rotations and the reflections multiplied by ${\cal U}$ is isomorphic
to $C_{4v}$. The $W=0$ pairs appropriate for half a flux quantum must 
involve two holes belonging to the degenerate irrep of $\tilde{C}_{4v}$.
In this way, at  $\phi=\phi_{0}/2$ the full symmetry is 
restored, allowing again for  pairing and negative $\Delta$.  
The $W=0$ quasiparticles have just the correct symmetry properties in the 
presence of the vector potential to provide superconducting flux 
quantisation in macroscopic systems.

$W=0$ pairs exist only in planar lattice structure 
with a certain point symmetry Group. So we expect that the CuO$_{2}$ geometry
is not the only one where $W=0$ property is present. In the following 
we want to show that in the one-band Hubbard Model too, where the 
point symmetry Group is still $C_{4v}$, we can build 
$W=0$ pairs and that they continue to exist even in the presence of 
magnetic field if and only if the magnetic flux is quantized in 
units of $\f_{0}/2$\cite{epj1999}.

Let us start with the one-body Hamiltonian defined on a square lattice 
with $N_{\L}\times N_{\L}$ sites\footnote{
Although 
we know that the actual cuprate superconductors present a band structure with three 
dominant bands, the one-band Hubbard Model tries to mimic the presence 
of the charge transfer gap $\D_{pd}$ by means of an {\it effective} value 
of the Coulomb repulsion and thus it presents only two bands. The 
``Oxygen'' band becomes the lower Hubbard band of this model.} 
(for the moment we don't 
care of the spin index), 
\begin{equation}
K=\frac{t}{2}\sum_{\bra\mathbf{r},\mathbf{r}'\ket}
(c^{\dag}_{\mathbf{r}}c_{\mathbf{r}'}+{\mathrm h.c.})+\e\sum_{\mathbf{r}}
c^{\dag}_{\mathbf{r}}c_{\mathbf{r}},\;\;\;\; 
 {\mathbf r}= (i_{x},i_{y}),\;\;\;i_{x},i_{y}=1,\ldots,N_{\L}
\label{kinobhm}
\end{equation}
where $\bra{\mathbf r},{\mathbf r}'\ket$ stands for nearest neighbors. 
We take the $x,y$ axes 
parallel to the bonds and fold the plane on itself after $N_{\L}$ 
lattice spacings in both directions, forming the torus shown in Figure 
\ref{magnfinplane} (B). 
Since we are dealing with  complex fermion fields, the most general 
boundary conditions are 
\begin{equation}
c_{{\mathbf r}+\tau_{a}}=e^{i\alpha}c_{{\mathbf r}}\;\;\;\;\;\;\; 
c_{{\mathbf r}+\tau_{b}}=e^{i\beta}c_{{\mathbf r}}
\end{equation}
where $\tau_{a}$ and $\tau_{b}$ are the torus periods: 
$\tau_{a}=(N_{\L},0)$ and $\tau_{b}=(0,N_{\L})$. 
The fermion fields can then be expanded as
\begin{equation}
c_{{\mathbf r}}=\frac{1}{N_{\L}}\sum_{{\mathbf k}}
e^{i({\mathbf k}+{\mathbf t})\cdot {\mathbf r}}
c_{{\mathbf k}},\;\;\;
{\mathbf k}=\frac{2\pi}{N_{\L}}(i_{x},i_{y}),\;\;\;i_{x},i_{y}=1,\ldots,N_{\L}
\label{fourexpff}
\end{equation}
with ${\mathbf t}=(\alpha/N_{\L},\beta/N_{\L})$. Turning on a costant magnetic field 
${\mathbf B}=(B_{a},B_{b})$ parallel to the surface of the torus, 
the flux through a circular surface $\S_{a}$ ($\S_{b}$) 
of radius $r_{0}=N_{\L}/2\pi$ perpendicular 
to the $b$ ($a$) direction  will be
\begin{equation}
\phi_{\t}=\int_{\S_{\t}}{\mathbf B}\cdot {\mathbf n}\,d\sigma = 
\oint_{\partial \S _{\t}}{\mathbf A}\cdot d{\mathbf l}=2\pi r_{0} 
A_{\t},\;\;\;\;\t=a,b
\end{equation}
with ${\mathbf n}$ the surface versor. The above equation enables us to compute, 
using the Peierls 
prescription, how the presence of this particular constant magnetic 
field changes the original Hamiltonian $K$ of Eq.(\ref{kinobhm}). 
The final result is
\begin{equation}
K\rightarrow \frac{t}{2}\left[\sum_{\bra{\mathbf r},{\mathbf r}'\ket_{a}}
(e^{\frac{2\pi i}{N_{\L}}(\phi_{a}/\phi_{0})}
c^{\dag}_{{\mathbf r}}c_{{\mathbf r}'}+{\mathrm h.c.})+
\sum_{\bra{\mathbf r},{\mathbf r}'\ket_{b}}
(e^{\frac{2\pi i}{N_{\L}}(\phi_{b}/\phi_{0})}
c^{\dag}_{{\mathbf r}}c_{{\mathbf r}'}+{\mathrm h.c.})\right]+
\e\sum_{{\mathbf r}}
c^{\dag}_{{\mathbf r}}c_{{\mathbf r}}
\label{kgauge}
\end{equation}
where $\bra{\mathbf r},{\mathbf r}'\ket_{\t}$ 
are two nearest neighbor sites along the $\t=a,b$ 
direction. The spectrum of 
the new Hamiltonian is the same of the original one if
$\alpha\rightarrow\alpha+2\pi\phi_{a}/\phi_{0}$ and 
$\beta\rightarrow\beta+2\pi\phi_{b}/\phi_{0}$. 
In this sense we can say that a change in boundary conditions is the 
same of turning on gauge fields\cite{alvarez}. Of course the 
choice which physically corresponds to the unperturbed infinite plane 
is $\alpha=\beta=0$ (Born-von Karman boundary conditions).

Denoting with
\begin{equation}
e^{i\beta}\begin{array}{c} e^{i\alpha} \\ \begin{tabular}{|c|}
        \hline \;\;\;\\
        \hline
    \end{tabular}  \\ \phi_{a} 
\end{array} \phi_{b}
\end{equation}
the configuration for which the boundary conditions are $e^{i\alpha}$  
and $e^{i\beta}$ and the flux is $\phi_{a}$ and $\phi_{b}$ along the 
the $a$ and $b$ direction respectively, explicit calculations show 
that for the Hamiltonian (\ref{kgauge}),  $W=0$ pairs builded from 
determinantal states of the form in Eq.(\ref{detercoop}) of $^{1}A_{2}$, 
$^{1}B_{2}$ and $^{1}B_{1}$ symmetry exist if and only if one of the 
following configurations is realized
\begin{equation}
 e^{i\beta}\begin{array}{ccc} e^{i\alpha} \\\begin{tabular}{|c|}
        \hline \;\;\;\\
        \hline
    \end{tabular} \\ 
  (n_{\alpha}-\alpha/\pi)\f_{0}/2 \end{array}
 (n_{\beta}-\beta/\pi)\f_{0}/2
\label{conf}
\end{equation}	 
with $n_{\alpha}$ and $n_{\b}$ integers such that 
$(-1)^{n_{\alpha}}=(-1)^{n_{\beta}}$. 

From Eq.(\ref{conf})  we see that only two configurations allow for
$W=0$ pairs in absence of 
magnetic field; they correspond to choose boundary conditions both periodic 
or  antiperiodic along  two orthogonal directions. With such 
boundary conditions   $W=0$ pairs exist even in a magnetic
field if and only if 
the flux is quantised in multiples of  half the fundamental 
fluxon. For any other choice of boundary conditions, flux is still quantised in 
a similar manner, but we need a zero point flux in order to have $W=0$ 
pairs, so that they do not exist at zero magnetic field. 
If we assume  dressed $W=0$  
pairs to be  quasiparticles of the superconducting phase, then  
flux is  quantized in a superconducting way. In the next Section we will
prove rigorously that this is the case, by studying the role of $W=0$  
pairs in the many-body problem and showing how they become a 
genuine bound state by means of the electronic correlation.}

\subsection{Canonical Transformation}
\label{canonical}

{\small
In the Cooper theory, an effective interaction involving phonons is
introduced via an approximate canonical transformation. In the 
Hubbard Model, the holes can exchange particle-hole pairs, 
and more complex electronic excitations, rather than phonons, but one 
can approach the problem in a similar 
way\cite{epj1999}\cite{ssc1999}\cite{ijmp1999}. 

Suppose the system is in its ground state 
with Fermi energy $\e_{F}$ and
a couple of extra holes are added. We wish to show that by a
canonical transformation one obtains an effective Hamiltonian which describes
the propagation of a pair of {\em dressed} holes, and includes
{\em all} many-body effects. Including all the diagrams of a 
generalized RPA would lead to something like the FLEX approximation 
\cite{bickers} whose implications for the superconductivity in 
the three-band Hubbard Model have been explored recently in a series 
of papers\cite{esirg}. A related self-consistent and 
conserving T-matrix approximation has been proposed by Dahm and 
Tewordt\cite{dahm} for the excitation spectra in the $2d$ Hubbard Model; 
we  mention incidentally that recently diagrammatic methods have been 
successfully applied to the photoelectron spectra of the cuprates in other contexts 
too, like the spin-fermion model\cite{sps}. 

The exact many-body ground state with two added holes may be expanded in
terms of excitations over the vacuum (the non-interacting Fermi {\em sphere}) 
by a configuration interaction: 
\begin{equation}
|\Psi_{0}(N+2)\ket={\sum_{m}}a_{m}|m\ket+{\sum_{\alpha }}a_{\alpha }|\alpha 
\ket+{\
\sum_{\beta }}a_{\beta }|\beta \ket+....  
\label{psi0}
\end{equation}
here $m$ runs over pair states, $\alpha $ over 4-body states ($2$ holes and $1$
e-h pair), $\beta $ over 6-body ones ($2$ holes and $2$ e-h pairs), 
and so on\footnote{Many configurations contribute to the interacting 
ground state and we need a complete set to expand it exactly. 
However, as long as the set is complete, the expansion is exact and 
well defined only if the non-interacting ground state is 
non-degenerate. We assume this is the case in Eq.(\ref{psi0}). 
Besides, we stress that Eq.(\ref{psi0}) is configuration 
interaction, {\it not a perturbative expansion}.}.
To set up the Schr\"{o}dinger equation, we consider the effects of the
operators on the terms of $|\Psi _{0}(N+2)\ket$. We write:
\begin{equation}
K|m\ket=\e_{m}|m\ket,\;\;\;K|\alpha \ket=\e_{\alpha }|\alpha \ket,
\;\;\;K|\b \ket=\e_{\b }|\b \ket,...  
\label{h0m}
\end{equation}
and since $W_{\mathrm{Hubbard}}$ can create or destroy up to 2 e-h pairs,
\begin{equation}
W_{\mathrm{Hubbard}}|m\ket={\sum_{m^{\prime }}}W_{m^{\prime },m}
|m^{\prime }\ket+{\sum_{\alpha }} W_{\alpha ,m}
|\alpha \ket  
+{\ \sum_{\beta }}W_{\beta ,m}|\beta \ket.  
\label{wm}
\end{equation}
$W_{m^{\prime },m}$ does not 
contribute to the effective interaction for $W=0$ pairs in our model; however we keep it
for the sake of generality. It allows to introduce the effect of the 
exchange of phonons and
other quasiparticles that we are not considering. For clarity let us
first write the equations that include explicitly up to 6-body states; then
we have 
\begin{equation}
W_{\mathrm{Hubbard}}|\alpha \ket={\sum_{m}} W_{m,\alpha }|m\ket+
{\sum_{\alpha ^{\prime }}}W_{\alpha ^{\prime },\alpha } |\alpha
^{\prime }\ket 
+{\sum_{\beta }}W_{\beta ,\alpha }|\beta \ket  
\label{walfa}
\end{equation}
where scattering between 4-body states is allowed by the second term, and
\begin{equation}
W_{\mathrm{Hubbard}}|\beta \ket={\sum_{m^{\prime }}}
W_{m^{\prime
},\beta }\left| m^{\prime }\right\rangle +
{\sum_{\alpha }}W_{\alpha ,\beta } \left| \alpha \right\rangle 
+{\sum_{\beta^{\prime } }}W_{
\beta^{\prime } ,\beta }\left|\beta^{\prime } \right\rangle  .
\label{wbeta}
\end{equation}
In principle, the $W_{\beta^{\prime } ,\beta }$ term can be eliminated by
taking linear combinations of the complete set of $\beta $ states such 
that $(K+W_{\mathrm{Hubbard}})_{\beta^{\prime } ,\beta }=\e'_{\beta }
\d_{\b\b'}$: when
this is done, we get a ``self-energy'' correction to $\e_{\beta }$ and a
mixing of the vertices, without altering the structure of the
equations. The Schr\"{o}dinger equation yields equations for the
coefficients $a_{m}$, $a_{\a}$ and $a_{\b}$ 
\begin{equation}
\left( \e_{m}-E_{0}\right) a_{m} 
+{\sum_{m^{\prime }}}W_{m,m^{\prime }}a_{m^{\prime }}+{\sum_{\alpha }}
W_{m,\alpha }a_{\alpha }+{\sum_{\beta }}W_{m,\beta } a_{\beta }=0
\label{eq65}
\end{equation}
\begin{equation}
\left( \e_{\alpha }-E_{0}\right) a_{\alpha }  
+{\sum_{m^{\prime }}}W_{\alpha,m^{\prime }}a_{m^{\prime }}+{\sum_{\alpha
^{\prime }}}W_{\alpha ,\alpha ^{\prime }}a_{\alpha ^{\prime }}+{\sum_{\beta }
}W_{\alpha ,\beta }a_{\beta } =0
\label{eq66}
\end{equation}
\begin{equation}
\left( \e'_{\beta }-E_{0}\right) a_{\beta }+{\sum_{m^{\prime }}}
W_{\beta ,m^{\prime }}a_{m^{\prime
}}+{\sum_{\alpha ^{\prime }}}W_{\beta ,\alpha ^{\prime }}a_{\alpha ^{\prime
}}=0  
\label{16}
\end{equation}
where $E_{0}$ is the interacting ground state energy. Then, we exactly decouple the
6-body states by solving the equation for $a_{\beta }$ and substituting 
into Eqs.(\ref{eq65})(\ref{eq66}), getting:
\begin{equation}
\left( \e_{m}-E_{0}\right) a_{m}+{\sum_{m^{\prime }}}\left[
W_{m,m^{\prime }}+{\sum_{\beta }}\frac{W_{m,\beta }W_{\beta ,m^{\prime }}}{
E_{0}-\e'_{\beta }}\right] a_{m^{\prime }}
+{\sum_{\alpha }}\left[ W_{m,\alpha }+{\sum_{\beta }}\frac{
W_{m,\beta }W_{\beta ,\alpha }}{E_{0}-\e'_{\beta }}\right]a_{\alpha } =0 
\label{17}
\end{equation}
\begin{equation}
\left( \e_{\alpha }-E_{0}\right) a_{\alpha }+{\sum_{m^{\prime }}}
\left[ W_{\alpha,m^{\prime }}+{\sum_{\beta }}\frac{W_{\alpha,
\beta }W_{\beta ,m ^{\prime }}}{E_{0}-\e'_{\beta }}\right]  a_{m^{\prime}}
+{\sum_{\alpha ^{\prime }}}\left[ W_{\alpha ,\alpha
^{\prime }}+{\sum_{\beta }}\frac{W_{\alpha
,\beta }W_{\beta ,\alpha ^{\prime }}}{E_{0}-\e'_{\beta }}\right]
a_{\alpha ^{\prime }}=0 
\label{18}
\end{equation}
Thus we see that the r\^{o}le of 6-body states is just to renormalize the
interaction between 2-body and 4-body ones, and for the rest they may be
forgotten. If $E_{0}$ is outside the continuum of excitations, as we
shall show in the next Section, the corrections are finite, and experience with clusters
suggests that they are small. Had we included 8-body excitations, we could
have eliminated them by solving the system for their coefficients and
substituting, thus reducing to the above problem with further
renormalizations. In principle, the method applies to all the higher order
interactions, and we can recast our problem as if only 2 and 4-body states
existed. Again, the $W_{\alpha^{\prime } ,\alpha }$ term can be eliminated
by taking linear combinations of the $\alpha $ states: when this is done,
we get a ``self-energy'' correction to $\e_{\alpha }$ and a mixing of
the $W_{m,\alpha }$ vertices. Denoting $W^{\prime}$ and $\e^{\prime}$ 
the renormalised quantities, which can be determined by the procedure 
outlined above, one can solve for $a_{\alpha }$. 
The eigenvalue problem becomes
\begin{equation}
\left( E_{0}-\e_{m}\right) a_{m}=\sum_{m^{\prime}}\left\{
W_{m,m^{\prime }}^{\prime }+\left\langle m|S[E_{0}]|m^{\prime }\right\rangle \right\}
 a_{m^{\prime }}, 
\label{schro}
\end{equation}
where
\begin{equation}
\left\langle m|S\left[ E_{0}\right] |m^{\prime }\right\rangle ={\sum_{\alpha
}}\frac{\bra m|W^{\prime }|\alpha \ket\bra\alpha |W^{\prime }
|m^{\prime }\ket}{E_{0}-\e_{\alpha }^{\prime }}.  
\label{eiv}
\end{equation}

Eq.(\ref{schro}) is of the form of a Schr\"{o}dinger equation with eigenvalue
$E_{0}$ for pairs with an effective interaction $W'+S$. Then we interpret
$a_{m}$ as the wave function of the dressed pair, which is acted upon by an
effective Hamiltonian $\tilde{H}_{\mathrm{pair}}$. This way of 
looking at Eq.(\ref{schro}) is perfectly consistent, despite the 
presence of the many-body eigenvalue $E_{0}$, because we are not 
compelled to reference the energy eigenvalues to the hole vacuum.  
We note that if we shift the kinetic operator 
$K$ by an arbitrary constant $\Delta E$ in Eq.(\ref{h0m}), by setting
$K'=K-\Delta E$, the same shift applies to the 
eigenvalues $\e_{m},\e_{\alpha },\e_{\beta }$ and so on, and also to the 
renormalised quantities like $\e^{\prime}_{\alpha 
},\e^{\prime}_{\beta }$. Therefore, the  matrix elements of the scattering 
operator $S$ are unaffected 
by the shift. Thus we can reference $E_{0}$ to a new energy origin 
by shifting the diagonal terms in Eq.(\ref{schro}) without changing 
the off-diagonal terms. Since we wish to regard Eq.(\ref{schro}) 
as a Cooper-like equation for the pair, it is natural to set $\Delta E$ 
equal to the interacting ground state energy eigenvalue for the 
$N$ hole system, relative to the hole vacuum. 
The energy of two independent holes, relative to the  $N$ 
background, is $2\e_{F}$; when the 
effective interaction is accounted for, the energy of the two bound 
holes is $2\e_{F}+\Delta$, where $|\Delta|$ is the binding energy. 

The change from the full 
many-body $H_{\mathrm{Hubbard}}$ to $
\tilde{H}_{\mathrm{pair}}$ is the canonical transformation to dressed 
pairs we were looking for. In the new picture, the holes interact through an
effective vertex with infinitely many contributions. 
The linked contributions represent repeated exchange of electron-hole
pairs, and may contain self-energy insertions; all these contributions make up
the effective interaction. The unlinked diagrams are pure self-energy. Thus, the
scattering operator $S$ is of the form
$S=W_{\mathrm{eff}}+F$, where
$W_{\mathrm{eff}}$ is the effective interaction between dressed holes, while $F$ is a
forward scattering operator, diagonal in the pair indices $m$ ,$m^{\prime }$ which
accounts for the self-energy corrections of the one-body propagators: it is
evident from Eq.(\ref{schro}) that it just redefines the dispersion law 
$\e_{m}$ and we suppose it be included in the kinetic 
term $K$. 

So, letting $|a\ket=\sum_{m}a_{m}\left| m\right\rangle$, 
the effective Schr\"{o}dinger equation for the pair reads
\begin{equation}
\left( K+W^{\prime }+W_{\mathrm{eff}}\right) |a\ket=E_{0}|a\ket 
\label{es}
\end{equation}
and pairing takes place if $E_{0}=2\e_{F}-| \Delta|
$; then  $| \Delta | $ is the binding energy of the
pair. Any other pairing mechanism not considered here, like off-site
interactions, inter-planar coupling and phonons, can be included as 
an extra contribution to $W_{m^{\prime },m}^{\prime }$ which  just adds to 
$W_{\mathrm{eff}}$.

We emphasized the fact that in principle the canonical transformation is exact because in
this way the framework does not require $U/t$ to be small. However to test 
the instability of the Fermi liquid towards pairing it is 
sufficient to study the amplitudes $a_{m}$ of the $m$ states. In the 
weak coupling limit this can be done truncating the expansion 
(\ref{psi0}) 
to the $\a$ states because, as we have shown, the inclusion 
of the $\b,\;\g,\ldots$ states 
produces a renormalization of the matrix elements of higher order in 
$U$, leaving the structure of the equations unaltered\footnote{
We want to stress that the truncated expansion of $|\Q_{0}(N+2)\ket$ in 
Eq.(\ref{psi0}) doesn't give a good approximation of the interacting ground state 
wave function but only of its weak 
coupling $a_{m}$ amplitudes.
Similarly in the BCS Model, 
from the Cooper equation (obtained by truncating $|\Q_{0}(N+2)\ket$ to the $m$ 
states) we can estimate only the pair coefficients of the ground state 
and not its full structure. Nevertheless this is enough to study 
bound states formation; indeed the energy gap 
of the pair in the Cooper theory and in the 
many-body BCS theory are identical. Moreover, the pair wave function of 
the Cooper Model resembles the one of the many-body BCS Model\cite{richai}.}. 

Let $|\F_{0}(N)\ket$ be the non-interacting ground state formed by 
filling all the energy levels up to $\e_{F}(N)$ for both spin up and 
down. If the added pair has vanishing total momentum, like the one in 
Eq.(\ref{detercoop}), the $\alpha $ states are
\begin{equation}
|\alpha \ket=c^{\dag}_{{\mathbf k}'+{\mathbf q}+{\mathbf k}_{2},\ua}
c_{{\mathbf k}_{2},\da}c^{\dag}_{-{\mathbf q},\da}c^{\dag}_{-{\mathbf 
k}',\da}|\F_{0}(N)\ket
\label{alfas}
\end{equation}
where $c_{{\mathbf k}_{2},\da}$ creates an electron state; 
hence $\e_{{\mathbf k}_{2}}<\e_{F}<\e_{{\mathbf k}'+{\mathbf q}+{\mathbf 
k}_{2}}$, $\e_{{\mathbf q}}$, $\e_{{\mathbf k}'}$. The 
$\alpha$ states with opposite spin indices arise from the specular diagram,
contribute similarly and yield a factor of 2 at the end. 
The interaction matrix element is
\beq
\bra \F_{0}(N)|c_{-{\mathbf k}',\da}c_{-{\mathbf q},\da}
c^{\dag}_{{\mathbf k}_{2},\da}c_{{\mathbf k}'+{\mathbf q}+{\mathbf k}_{2},\ua}
W_{\mathrm{Hubbard}}|d[{\mathbf k}]\ket=\nonumber \\ 
\d({\mathbf q}-{\mathbf k})U({\mathbf k}'+{\mathbf q}+{\mathbf k}_{2},
-{\mathbf k}',{\mathbf k},{\mathbf k}_{2})-
\d({\mathbf k}'-{\mathbf k})U({\mathbf k}'+{\mathbf q}+{\mathbf 
k}_{2},-{\mathbf q},{\mathbf k},{\mathbf k}_{2})
\label{me}
\eeq
where the $U$ matrix is computed on the Bloch basis, see Section 
\ref{bands}. This expression holds for any lattice with translational 
invariance. Working out Eq.(\ref{eiv}) we find that 
the product in the numerator  yields 4
terms; two are proportional to $\delta ({\mathbf p}-{\mathbf q})$ 
and belong to $F$, while the cross
terms yield identical contributions to $W_{\mathrm{eff}}$. Thus we obtain the 
following effective interaction between the members of $W=0$ pairs:
\begin{eqnarray}
\left\langle \Phi_{\eta}\left[ {\mathbf p}\right] \right|W_{\mathrm{eff}}\left|
\Phi_{\eta}\left[ {\mathbf q}\right] \right\rangle =4\sum_{R\in 
C_{4v}}\chi^{(\eta)}(R)
\sum_{{\mathbf k}\in D(\e_{F})}
\theta\left( \e_{{\mathbf p} +{\mathbf q}_{R} +{\mathbf k}} -\e_{F}\right)
\times \nonumber\\ \times \frac{
U\left({\mathbf p} +{\mathbf q}_{R} +{\mathbf k},-{\mathbf p},{\mathbf q}_{R},
{\mathbf k}\right) U\left( {\mathbf p},{\mathbf k},
{\mathbf p} +{\mathbf q}_{R} +{\mathbf k},-{\mathbf q}_{R}\right)}
{\e_{{\mathbf p} +{\mathbf q}_{R} +{\mathbf k}}-\e_{{\mathbf k}}+
\e_{{\mathbf q}}+\e_{{\mathbf p}}-E_{0}},
\label{weffective}
\end{eqnarray}
while it can be shown that the corresponding matrix elements of the forward scattering 
operator $F$ vanish identically as in the fully symmetric 
clusters\cite{gimelli}.  
The sum in Eq.(\ref{weffective}) is over occupied ${\mathbf k}$ with empty 
${\mathbf p} +{\mathbf q}_{R} +{\mathbf k}$. Note that $W_{\mathrm{eff}}$
does not depend on the sign of $U$. 
The diagonal elements of Eq.(\ref{weffective}) are 
also clearly related to the $\Delta^{\left(2 \right)} $ expression 
(\ref{deltaiiord}) derived from
perturbation theory for the fully symmetric clusters\cite{cb5}, 
which are special cases of the present theory. The expressions 
(\ref{deltaiiord}) and (\ref{weffective}) are characterized 
by the symmetry-induced quantum mechanical interference of several terms. 
The sum can be positive 
or negative depending on the irrep. 
This interference produces a partial cancellation, and the absolute value of the 
result is typically much smaller than individual contributions. This 
explains why the interaction is dynamically small for $W=0$ pairs:  
they  have no direct interaction, and because of the interference the effective interaction is 
reduced compared to what one could expect by a rough 
order-of-magnitude estimate. However, the presence of the $\theta$ 
functions and the anisotropy of the integrands prevent a total 
cancellation. 

In Eq.(\ref{deltaiiord}), the vanishing 
of the denominators is prevented by the discrete spectrum of the 
cluster. In the thermodynamic limit this 
complication disappears in the full expression (\ref{weffective}) 
since we are  interested in bound pairs; then, $E_{0}$ is below the 
continuum and the denominators never vanish. 

The ${\mathbf p}$ and ${\mathbf q}$ indices run over $1/8$ of the 
empty BZ for $W=0$ pairs. We denote
such a set of empty states $D(\e_{F})/8$, and cast the result in the form of a
(Cooper-like) Schr\"{o}dinger equation
\begin{equation}
2\e_{{\mathbf k}} a_{{\mathbf k}} +
\sum_{{\mathbf k}'\in D(\e_{F})/8}W_{\mathrm{eff}}
\left( {\mathbf k},{\mathbf k}^{\prime }\right) a_{{\mathbf k}'} =
E_{0}a_{{\mathbf k}}  
\label{equazione}
\end{equation}
for a self-consistent calculation of $E_{0}$ (since $W_{\mathrm{eff}}$ depends on the
solution). }

\subsection{Pairing in the CuO$_{2}$ Plane: Numerical Results}
\label{numerical}

{\small 
Using the analytical expression for the 
effective interaction in the  full plane, Eq.(\ref{weffective}), we have performed 
exploratory numerical estimates of $\Delta$ by working on supercells of 
$N_{\L}\times N_{\L}=|\L|/3$ cells, with periodic boundary 
conditions. For the sake of
simplicity, we shall neglect the minor contributions from the higher bands and
consider the dominant intra-band processes, in which empty states belong to the
bonding band. Here we 
solved the problem in a virtually exact way for $N_{\L}$ up to 40.
Several good supercell calculations devoted to the problem of pairing  
have been reported to date\cite{dagrev}, but no conclusive 
evidence was reached, because of 
the difficulty of dealing with size effects. First, we searched for 
triplet solutions with negative energy without success, since, as in the
clusters, $W_{\mathrm{eff}}$ is repulsive for triplets. On the 
contrary, $W=0$  
singlets did show pairing, in line with our previous findings in
small clusters\cite{cb1}\cite{cb2}\cite{cb34}\cite{cb5}.
We took as input data the set of current parameters, already used for 
clusters, that is  (in eV) $t_{pd}=1.3$, $\e_{p}=3.5$, 
$\e_{d}=0$, $U_{p}=6s$, $U_{d}=5.3s$, where $s$ is a scale
factor induced by renormalization. Since screening excitations are 
explicitly accounted for in the Hamiltonian, it is likely that $U$ is a 
bare (unscreened) quantity, which justifies $s>1$. A stronger 
interaction causes smaller pairs and speeds up convergence within 
attainable supercell sizes. In Table \ref{epj99II}, we report the results for 
$^{1}B_{2}$ pairs at $s=\frac{3}{\sqrt{2}}=2.121$ with 
$\e_{F}=-1.3$ eV (half filling corresponds to $\e_{F}=-1.384$ eV).
We see that although $|\Delta|$ decreases monotonically with increasing
supercell size, $V_{\mathrm{eff}}$, see below for its definition, 
is {\em not} dropping to zero, but remains fairly stable around 6 to 7 eV. 
\begin{table}[tbp]
    \centering
    \caption{{\footnotesize Binding energy of $^{1}B_{2}$ pairs in 
    supercells. }}
   \begin{tabular}{|c|c|c|c|c|}
       \hline
$N_{\L}$ & $ \r=N/N^{2}_{\L}$ & $ -\Delta$ (meV) & 
$V_{\mathrm{eff}}$ (eV)  & $-\Delta _{\mathrm{asympt}}$ (meV)   \\ 
 \hline
18 & 1.13 & 121.9 & 7.8 & 41.6   \\ 
 \hline
20 & 1.16 & 42.2 & 5.0 & 9.0   \\ 
 \hline
24 & 1.14 & 59.7 & 7.0 & 28.9   \\ 
 \hline
30 & 1.14 & 56. & 5.7 & 13.2   \\ 
 \hline
40 & 1.16 & 30.5 & 6.6 & 23.4 \\
 \hline
\end{tabular}
\label{epj99II}
\end{table}
\begin{table}[tbp]
    \centering
    \caption{{\footnotesize Effective Cooper parameters for fitting the 
    $V_{\mathrm{eff}}$ dependence of $\Delta$ at several $\e_{F}$ values.}}
\begin{tabular}{|c|c|c|}
    \hline
$\e_{F}$ (eV) & $\omega_{D}$ (eV) & $\rho_{F}$ (eV$^{-1}$/cell)  \\ 
 \hline
-1.35 & 0.4545 & 0.0486   \\ 
 \hline
-1.3 & 0.492 & 0.0415   \\ 
 \hline
-1.2 & 0.520 &0.0338   \\ 
 \hline
-1.1 & 0.5237 & 0.02905   \\ 
 \hline
\end{tabular}
\label{epj99III}
\end{table}
With supercell sizes $N_{\L}>40$ calculations become hard. Since we 
are concerned with the asymptotic behavior for
$N_{\L}\rightarrow\infty $ and  $\Delta$ depends on $U$'s and $N_{\L}$ in
a peculiar way, increasing with 
$N_{\L}$ for large $U$'s and dropping for small $U$'s, 
we need a simple solvable model in supercells and in the 
infinite plane to make reliable extrapolations of numerical results.
To this end, we define the Uniform Interaction Model (UIM) in 
which a constant negative interaction $-V_{\mathrm{eff}},\;V_{\mathrm{eff}}>0$
prevails for $\mathbf{k}$ and $\mathbf{k}^{\prime }$ in $D(\e_{F})/8$. In the 
Cooper formula for the binding energy of the pair one has 
\begin{equation}
|\Delta|=\frac{2 \omega_{D}}{e^{\frac{1}{\g\rho_{F}}}-1},
\label{coop}
\end{equation}
where $\rho_{F}$ is the density of states at the Fermi level and  
$-\g$ is a (constant) attractive interaction per cell in a shell of 
thickness $2 \omega_{D}$ surrounding the Fermi surface 
(the orders of magnitude are $\omega_{D}\approx 25$  meV, 
$\rho_{F}\approx 0.3$ eV$^{-1}$/cell and $\g\approx 1$ eV/cell, 
and $|\Delta|$ is in the meV range); on the other 
hand, in the UIM no Debye frequency  $\omega_{D}$ is involved, and all the empty 
states in  $D(\e_{F})/8$ contribute to $\Delta$. Both the Cooper and the UIM 
models can be solved in supercell calculations, and show the same qualitative
behavior with increasing $N_{\L}$.
\begin{table}[tbp]
    \centering
    \caption{{\footnotesize Data for $^{1}B_{2}$ pairs at 
    $\e_{F}=-1.2$ eV; $V_{\mathrm{eff}}$ is in eV.}}
\begin{tabular}{lclclclclclclcl}
    \hline
 & $s=$ & $\sqrt{2}$ & $s=$ & $\frac{3}{\sqrt{2}}$ & $s=$ & 
 $2\sqrt{2}$ \\
 \hline 
$N_{\L}$ & $-\Delta$ (meV) & $V_{\mathrm{eff}}$ & $-\Delta$ (meV) & 
$V_{\mathrm{eff}}$ & $-\Delta$ (meV) & $V_{\mathrm{eff}}$ \\
\hline 
  12 &126 &6.75& 251.&10.&469.&14.23  \\
  \hline 
  20 & 63. &6.9 &135.&10.6&320.5&15.9  \\ 
  \hline 
30 & 27. &6.3& 198. &13.8&460.&20.8   \\ 
\hline 
$\infty$ & 8.8 &6.3& 131.&13.8& 333.&20.8  \\
\hline 
\end{tabular}
\label{epj99IV}
\end{table}
\begin{table}[tbp]
    \centering
    \caption{{\footnotesize Data for $^{1}B_{2}$ pairs at 
    $\e_{F}=-1.1$ eV; $V_{\mathrm{eff}}$ is in eV.}}
\begin{tabular}{lclclclclclclcl}
    \hline
&$s=$&$\sqrt{2}$&$s=$&$\frac{3}{\sqrt{2}}$&$s=$&$2\sqrt{2}$\\
\hline
$N_{\L}$ & $-\Delta$ (meV) & $V_{\mathrm{eff}}$& $-\Delta$ (meV) &
$V_{\mathrm{eff}} $& $-\Delta$ (meV) & $V_{\mathrm{eff}}$ \\
\hline
  12 &163. &9.& 409.&15.7&688.&22.1  \\
  \hline
  20 & 151. &10.8 &371.&17.6&635.5&24.3  \\ 
  \hline
30 & 80. &10.3& 270. &17.66&520.&24.9  \\ 
\hline
$\infty$ & 35.6 &10.3& 170.8&17.66& 362.&24.9 \\
\hline
\end{tabular}
\label{epj99V}
\end{table}

For small $N_{\L}$, $\Delta$  is of the same
order as $V_{\mathrm{eff}}$, while in the thermodynamic limit ($N_{\L}\rightarrow 
\infty$), the asymptotic value of $\Delta$, $\Delta_{\mathrm{asympt}}$, can 
easily be estimated. Setting 
$W_{\mathrm{eff}}=-V_{\mathrm{eff}}/N^{2}_{\L}$, we get from 
Eq.(\ref{equazione})
\begin{equation}
(2\e_{\mathbf{k}}-E_{0})a_{\mathbf{k}}=\frac{V_{\mathrm{eff}}}{8}
\frac{1}{(2\pi)^{2}}\int d{\mathbf k}'\,a_{\mathbf{k}'}=\mathrm{const}.
\end{equation}
Writing $a_{\mathbf{k}}={\mathrm const}/(2\e_{\mathbf{k}}-E_{0})$ and 
$E_{0}=2\e_{F}-|\Delta|$, we obtain
\begin{equation}
\frac{8}{V_{\mathrm{eff}}}=\int_{\e_{F}}^{0}\frac{d\epsilon \rho \left( \e\right) 
}{2\left( \e -\e_{F}\right) +|\Delta_{\mathrm{asympt}}|}  .
\label{32}
\end{equation}
Here, $\rho$ is not a constant, as in the Cooper formula, and its 
values in the integration range are sensitive to filling; howewer its 
order of magnitude does not differ much from the BCS case except at 
half filling. Good fits can be obtain using the Cooper formula
with effective $\omega_{D}$ and $\rho_{F}$ parameters, like 
those shown in Table \ref{epj99III}.

In Table \ref{epj99IV}, we consider $^{1}B_{2}$ pairs at $\e_{F}=-1.2$ eV and
report values of $\Delta$
and $V_{\mathrm{eff}}$ for supercell calculations at $N_{\L}$=12, 20 and 30
for 3 values of the scale factor $s$. The filling is 
$\r \sim 1.3$ at
$N_{\L}=30$. Fixing $V_{\mathrm{eff}}$ at the value of $N_{\L}=30$ we 
calculate $\Delta_{\mathrm{asympt}}$ for $N_{\L}\rightarrow\infty$. 

We see that the relatively mild $N_{\L}$ dependence of 
$V_{\mathrm{eff}}$ supports the use
of the UIM to extrapolate the results to the thermodynamic limit, and there is a
clear indication of pairing with sizable binding energies. The $s$ dependence of 
$V_{\mathrm{eff}}$ is roughly linear, while $\Delta$ depends exponentially on $s$.
Table \ref{epj99V} presents the results for  $^{1}B_{2}$ pairs at $\e_{F}=-1.1$ eV 
($\r\approx 1.4$ at $N_{\L}=30$). The trend is similar, but $V_{\mathrm{eff}}$ is seen
to increase with doping.

The results for the $^{1}A_{2}$ pairs are seen to
lead to bound states as well, with comparable $\Delta$ 
values\cite{epj1999}; the trend with
doping is opposite, however, and the gap is nearly closing at 
$\e_{F}=-1.1$ eV.
A necessary condition for superconducting flux quantization is that two kinds of
pairs of similar binding energy and different symmetries 
exist\cite{cbs1}. A similar conclusion was reached independently by 
other authors\cite{evert}. Moreover, evidence of mixed $(s+id)$ 
symmetry for the pairing state has been amply reported in 
angle-resolved photoemission studies \cite{shen}. 
This remark leads to the prediction that in
this model superconductive pairing disappears with increasing $\r$, while
a different sort of pairing prevails; in reality the CuO$_{2}$ plane
prefers to distort at excessive doping, 
and in a distorted plane the present mechanism,
based on symmetry, could be destroyed.}

\subsection{Antiferromagnetic Ground State at Half Filling}
\label{afgs}

{\small
Up to now, we learned how to master the situation when the number $N$  of holes
in the 
$N_{\L}\times N_{\L}$ system is such that the non-interacting ground 
state $|\Phi_{0}(N)\ket$ is a single 
non-degenerate determinant (the Fermi surface is totally filled). Then, 
switching the repulsion $W_{\mathrm{Hubbard}}$ on, 
we know how to calculate the effective 
interaction $W_{\mathrm{eff}}$ 
between  two {\em extra} holes added to the 
system by the expansion (\ref{psi0}).  However we are particularly interested in 
the doped antiferromagnet, and the antiferromagnetic ground state 
occurs at half filling, {\em not} in a filled-shell situation.

We want to study the doped antiferromagnet since there are strong indications
that the Fermi liquid is unstable towards pairing  near half filling; 
they come from diagrammatic 
approaches\cite{bickers}, renormalization group 
techniques\cite{zanchi}\cite{metzner} and also cluster diagonalizations\cite{cb5}
\cite{fop}\cite{fettes}. Therefore exact results on the half filled 
Hubbard Model may be relevant to antiferromagnetism  
and to the mechanism of the superconducting instability 
as well.  On the other hand, pairing by the ``$W=0$'' mechanism does 
{\em not} require that the pair be created on a single-determinant state.

For simplicity in the rest of this paper we concentrate on the one-band Hubbard 
Model, defined by Eq.(\ref{kinobhm}) plus the usual on-site Hubbard 
interaction $W_{\mathrm{Hubbard}}$: 
\begin{equation}
H_{\mathrm{Hubbard}}=K+W_{\mathrm{Hubbard}}=
\frac{t}{2}\sum_{\bra{\mathbf r},{\mathbf r}'\ket\s}
(c^{\dag}_{{\mathbf r}\s}c_{{\mathbf r}'\s}+{\mathrm h.c.})+U\sum_{{\mathbf r}}
\hat{n}_{{\mathbf r}\ua}\hat{n}_{{\mathbf r}\da}
\label{hhafgs}
\end{equation}
with ${\mathbf r}=(i_{x},i_{y})$, $i_{x},i_{y}=1,\ldots,N_{\L}$, 
the sum on $\bra{\mathbf r},{\mathbf r}'\ket$ is over 
the pairs of nearest neighbors sites and 
$\hat{n}_{{\mathbf r}\s}$  is the number operator on the site 
${\mathbf r}$ of spin $\s$\footnote{Here we have dropped the one-body 
on-site potential $\e\sum_{{\mathbf r}\s}
c^{\dag}_{{\mathbf r}\s}c_{{\mathbf r}\s}$, since we work with a fixed 
number of spin up and down particles.}. The point symmetry is 
$C_{4v}$; besides, $H_{\mathrm{Hubbard}}$ is 
invariant under the  commutative Group of translations ${\mathbf 
T}$ and hence the space Group\cite{hamer}   ${\mathbf G}={\mathbf 
T} \otimes C_{4v} $; $\otimes$ means the semidirect product. 
In terms of the Fourier expanded fermion operators (periodic boundary 
conditions) 
$c_{{\mathbf k}}=\frac{1}{N_{\L}}\sum_{{\mathbf r}}e^{i 
{\mathbf k}\cdot {\mathbf r}}c_{{\mathbf r}}$, we have 
$K=\sum_{{\mathbf k}}\e_{{\mathbf k}}c^{\dag}_{{\mathbf k}\s}c_{{\mathbf k}\s}$ 
with $\e_{{\mathbf k}}=2t[\cos k_{x}+\cos k_{y}]$. 
Then, the one-body plane wave state $c^{\dag}_{{\mathbf k}\s}|0\ket
\equiv|{\mathbf k}\s\ket$ is an  eigenstate of $K$.

The $4\times 4$ cluster is half filled at $N=16$. Below, we shall consider 
its ground state  with $N=4$\cite{ijmp00} and $N=14$\cite{fop}\cite{epj01} 
holes. For $U \rightarrow 0$, $N=4$ means that 2 holes are added to a 
closed shell, while  $N=14$ corresponds to 2 electrons added to the 
half filled system.  We demonstrate that in both cases the structure of the ground 
state must be interpreted in terms of 
pairing\footnote{Hole pairing in the former case  and electron pairing in the latter.}.
However the non-interacting ground 
state at half filling is degenerate and in order to deal with $N=14$  we must 
extend the above formalism. 

As a preliminary measure in this Section we build the exact ground state of the Hubbard 
Model at half filling  and weak coupling for a general {\it even} 
$N_{\L}$\cite{ssc2001}\cite{jop2001}. 
Once a unique non-interacting ground state is determined, one can use 
the non-perturbative canonical transformation to test the instability 
of the system towards pairing; this will be done in the next Section 
for $N_{\L}=4$.

The starting point is the following property of the number operator 
$\hat{n}_{{\mathbf r}}=c^{\dag}_{{\mathbf r}}c_{{\mathbf r}}$ (for the moment we omit
the spin index).

\underline{{\em Theorem}}:   Let ${\cal S}$ be an arbitrary set 
of plane-wave eigenstates $\{|{\mathbf k}_{i}\ket\}$ of $K$ and
$(n_{{\mathbf r}})_{ij}=\bra {\mathbf k}_{i}|\hat{n}_{{\mathbf r}}|
{\mathbf k}_{j}\ket=\frac{1}{N_{\L}^{2}}
e^{i({\mathbf k}_{i}-{\mathbf k}_{j})\cdot{\mathbf r}}$  the matrix of 
$\hat{n}_{{\mathbf r}}$ in ${\cal S}$. 
This  matrix  has 
eigenvalues $\l_{1}=\frac{|{\cal S}|}{N_{\L}^{2}}$  and
$\l_{2}= \ldots =\l_{|{\cal S}|}=0$. 

Note that  $|{\cal S}|\leq N_{\L}^{2}$; if $|{\cal S}|= N_{\L}^{2}$ 
the set is complete, like the  set of all orbitals, and the theorem is 
trivial (a particle on site ${\mathbf r}$ is the 
$\hat{n}_{{\mathbf r}}$ eigenstate with eigenvalue 
1). Otherwise, if $|{\cal S}|< N_{\L}^{2}$, the theorem is an immediate 
consequence of the fact\cite{jop2001} that
\begin{equation}
    {\mathrm det}|(n_{{\mathbf r}})_{ij}-\l\d_{ij}|=(-\l)^{|{\cal S}|-1}
    (\frac{|{\cal S}|}{N_{\L}^{2}}-\l),\;\;\;\forall {\mathbf r}.
    \label{det}
\end{equation}

Let ${\cal S}_{hf}$ denote the set (or shell) of the ${\mathbf k}$ wave vectors 
such that $\e_{{\mathbf k}}=0$. 
At half filling ($N^{2}_{\L}$ particles) for $U=0$ the ${\cal S}_{hf}$  shell
is half occupied, while all ${\mathbf k}$ orbitals such that 
$\e_{{\mathbf k}}<0$ are filled. The ${\mathbf k}$ vectors of 
${\cal S}_{hf}$ lie on the square having 
vertices $(\pm\pi,0)$ and $(0,\pm\pi)$;
one  readily realizes that the dimension of the 
set ${\cal S}_{hf}$ is $|{\cal S}_{hf}|=2N_{\L}-2$. Since $N_{\L}$ is even 
and $H_{\mathrm{Hubbard}}$ commutes with the total spin operators, 
at half filling every ground state of $K$ is represented in 
the $S^{z}=0$ subspace. Thus, $K$ has  
$\left(\begin{array}{c} 2N_{\L}-2 \\ N_{\L}-1 
\end{array}\right)^{2}$   degenerate unperturbed ground state 
configurations with $S^{z}=0$. We wish to study below how this degeneracy is removed by the 
Hubbard interaction $W_{\mathrm{Hubbard}}$ already in first-order perturbation theory. 
Actually most of the degeneracy is removed in first-order, and  with the help of 
a theorem by Lieb\cite{lieb2th} we shall be  
able to single out the true, unique  ground state of $H_{\mathrm{Hubbard}}$ 
in Eq.(\ref{hhafgs})\footnote{
The Lieb theorem states that 
at half filling the ground state for a bipartite lattice is unique
and has spin 
$||{\cal S}_{1}|-|{\cal S}_{2}||/2$
where $|{\cal S}_{1}|$ 
($|{\cal S}_{2}|$) is the number of sites 
in the ${\cal S}_{1}$ (${\cal S}_{2}$) sublattice.}.
The first-order splitting of the degeneracy is obtained by diagonalizing the 
$W_{\mathrm{Hubbard}}$  matrix over the unperturbed basis; like in elementary 
atomic physics, the filled shells just produce a constant shift of all 
the eigenvalues and for the rest may be ignored in first-order 
calculations. In other terms we consider the {\it  truncated  Hilbert space ${\cal H}$} 
spanned by the {\it states of $N_{\L}-1$ holes of each spin in ${\cal 
S}_{hf}$}, and  we want the {\it exact} ground state(s) of $W_{\mathrm{Hubbard}}$ 
in  ${\cal H}$; by construction ${\cal H}$ is in the kernel of $K$, so the ground 
state of $W_{\mathrm{Hubbard}}$ is the ground state of $H_{\mathrm{Hubbard}}$ as well.
We call $W=0$ state any vector in ${\cal H}$ which also belongs to the kernel of 
$W_{\mathrm{Hubbard}}$.
Since the lowest eigenvalue of $W_{\mathrm{Hubbard}}$ is zero, 
it is evident that any $W=0$ state 
is a ground state of $H_{\mathrm{Hubbard}}$. We want to calculate the unique
ground state of the Hubbard Hamiltonian for $U=0^{+}$ at half
filling which
is a $W=0$ singlet with $2N_{\L}-2$ holes in ${\cal S}_{hf}$ (filled 
shells are understood)\footnote{Equivalently, we shall find the exact 
$(2N_{\L}-2)$-body ground state of the  
effective Hamiltonian 
\begin{equation}
H_{\mathrm{eff}}=\frac{U}{N_{\L}^{4}}\sum_{{\mathbf k}_{1},{\mathbf k}_{2},
{\mathbf k}_{3},{\mathbf k}_{4}\in {\cal S}_{hf}}
\d({\mathbf k}_{1}+{\mathbf k}_{2}-{\mathbf k}_{3}-{\mathbf k}_{4})
c^{\dag}_{{\mathbf k}_{1}\ua}c^{\dag}_{{\mathbf k}_{2}\da}
c_{{\mathbf k}_{3}\da}c_{{\mathbf k}_{4}\ua}
\end{equation}
where $\d(G)=1$ if $G$ is a reciprocal lattice vector and zero 
otherwise.}. 

To diagonalize the {\em local } operator $W_{\mathrm{Hubbard}}$ 
in closed form  we need to set up
a {\em local }  basis set of one-body states. If  ${\cal S}_{hf}$ were 
the 
complete set of plane-wave states 
${\mathbf k}$, the new basis would be trivially obtained by a Fourier 
transformation, but this is not the case. The question is: how can we 
define for each site ${\mathbf r}$ the  local counterparts of ${\mathbf k}$ states 
 using only those 
that belong to a degenerate level? 
The answer is: build  
a  set $\{|\varphi_{\a}^{({\mathbf r})}\ket\}$ of orbitals such that 
the number operator $\hat{n}_{{\mathbf r}}$ and the Dirac characters of the point symmetry 
Group $C_{4v}$ are diagonal. Using such a basis set for the 
half-filled shell the unique 
properties of the antiferromagnetic ground state become simple and 
transparent. 
The 
eigenvectors $|\varphi_{\a}^{(0)}\ket$ of $n_{{\mathbf r}=0}$ and those 
$|\varphi^{({\mathbf r})}_{\a}\ket$ of other sites ${\mathbf r}$ are connected by translation 
and also by a unitary transformation, or change of basis set. 
Picking ${\mathbf r}=\hat{e}_{l}$,  $l=x$ means $\hat{e}_{l}=(1,0)$ or transfer 
by one step towards the right and $l=y$ means $\hat{e}_{l}=(0,1)$ or
transfer by one step 
upwards.
The unitary transformation   reads:
\begin{equation}
|\varphi^{(\hat{e}_{l})}_{\a}\rangle=
\sum_{\b=1}^{2N_{\L}-2}|\varphi_{\b}^{(0)}\rangle
\bra\varphi_{\b}^{(0)}|\varphi^{(\hat{e}_{l})}_{\a}\rangle
\equiv \sum_{\b=1}^{2N_{\L}-2}|\varphi_{\b}^{(0)}\rangle T_{l_{\b\a}}.
\label{transferT}
\end{equation}
The transfer matrix $T_{l}$  {\em knows} all the translational and point symmetry of the 
system, and will turn out to be very special.

For large $N_{\L}$, to find $\{|\varphi_{\a}^{({\mathbf r})}\ket\}$ 
it is convenient to separate the ${\mathbf k}$'s of ${\cal S}_{hf}$ in 
irreducible representations of the space Group $\mathbf{G}$$=
C_{4v}\otimes {\mathbf T}$. Choosing an arbitrary ${\mathbf k}\in {\cal 
S}_{hf}$ with $k_{x}\geq k_{y}\geq 0$, the set of vectors $R_{i}{\mathbf k}$,  
where $R_{i}\in C_{4v}$, is a (translationally invariant) basis 
for an irrep of $\mathbf{G}$. The high
symmetry vectors $(0,\pi)$ and $(\pi,0)$ 
always trasform among themselves and are
the basis of the only two-dimensional irrep of $\mathbf{G}$, 
which exists for any $N_{\L}$.
If $N_{\L}/2$ is even,  one also finds the high symmetry  wavevectors 
${\mathbf k}=(\pm\pi/2,\pm\pi/2)$ which mix among themselves and yield
a four-dimensional irrep.  In general, the vectors $R_{i}{\mathbf k}$ are all 
different, so  all the other irreps of $\mathbf{G}$ have dimension 8, 
the number of operations of the point Group $C_{4v}$. 

Next, we show how to build our {\it local} basis set and derive $W=0$ 
states  for each kind of 
irreps of $\mathbf{G}$. For illustration, we will consider the 
case $N_{\L}=4$; then  
${\cal S}_{hf}$ contains the bases of two irreps of $\mathbf{G}$,
of dimensions 2 and 4. The one with basis
${\mathbf k}_{A}=(\p,0),\;{\mathbf k}_{B}=(0,\p)$ 
breaks into $A_{1}\oplus B_{1}$ in $C_{4v }$.

The eigenstates of $(n_{{\mathbf r}=0})_{ij}=\bra 
{\mathbf k}_{i}|\hat{n}_{{\mathbf r}=0}|{\mathbf k}_{j}\ket$, 
with $i,j=A,B$, 
are $|\q''_{A_{1}}\ket=\frac{1}{\sqrt{2}}(|{\mathbf k}_{A}\ket+|{\mathbf k}_{B}\ket)$ 
with $\lambda_{1}=1/8$
and $|\q''_{B_{1}}\ket=\frac{1}{\sqrt{2}}(|{\mathbf k}_{A}\ket-|{\mathbf k}_{B}\ket)$
with $\lambda_{2}=0$.
Since under translation by a lattice 
step $T_{l}$ along the $l=x,y$ direction 
$|{\mathbf k}\ket\ra e^{ik_{l}} |{\mathbf k}\ket$, using 
Eq.(\ref{transferT}) one finds that 
$|\q''_{A_{1}}\ket\leftrightarrow (-1)^{\th''_{l}}|\q''_{B_{1}}\ket$, 
with $\th''_{x}=1,\;\th''_{y}=0$;  so $|\q''_{A_{1}}\ket$ has 
vanishing amplitude on 
a sublattice and $|\q''_{B_{1}}\ket$ on the other. 
The two-body state $|\q''_{A_{1}}\ket_{\s}|\q''_{B_{1}}\ket_{-\s}$
has occupation for spin $\s$ but not for spin $-\s$ on the site ${\mathbf r}=0$; 
under a lattice  step 
translation it flips the spin and picks up a (-1) phase factor: 
$|\q''_{A_{1}}\ket_{\s}|\q''_{B_{1}}\ket_{-\s}
\leftrightarrow 
|\q''_{B_{1}}\ket_{\s}|\q''_{A_{1}}\ket_{-\s}$; therefore it has 
double occupation nowhere and is a 
$W=0$ state ($W=0$ pair \cite{ssc1999}\cite{epj1999}).

The 4-dimensional irrep with basis 
${\mathbf k}_{1}=(\p/2,\p/2),\;{\mathbf k}_{2}=(-\p/2,\p/2),
\;{\mathbf k}_{3}=(\p/2,-\p/2)\; 
{\mathbf k}_{4}=(-\p/2,-\p/2)$ breaks into $A_{1}\oplus B_{2}\oplus 
E$ in $C_{4v}$; letting $I=1,2,3,4$ for the 
irreps $A_{1},\;B_{2},\;E_{x},\;E_{y}$  
respectively, we can  write down all the eigenvectors 
of $(n_{{\mathbf r}=0})_{ij}=\bra {\mathbf k}_{i}|
\hat{n}_{{\mathbf r}=0}|{\mathbf k}_{j}\ket$,
with $i,j=1,\ldots,4$, as 
$|\q'_{I}\ket=\sum_{i=1}^{4}O'_{Ii}|{\mathbf k}_{i}\ket$, where $O'$ is the 
following 4$\times$4 orthogonal matrix
\begin{equation}
O'=\frac{1}{2}\left[\begin{array}{rrrr}
1 & 1 & 1 & 1 \\
1 & -1 & -1 & 1 \\
1 & -1 & 1 & -1 \\
-1 & -1 & 1 & 1 \end{array}\right].
\end{equation}
The state with non-vanishing eigenvalue is again the $A_{1}$ 
eigenstate. After a little bit of algebra we have shown\cite{jop2001} that
under 
$T_{l}$ the subspace of  $A_{1}$ and $B_{2}$ symmetry is 
exchanged with the one  of $E_{x}$ and $E_{y}$ symmetry. Thus 
we can build a 4-body eigenstate of $W_{\mathrm{Hubbard}}$ with vanishing eigenvalue: 
$|\q'_{A_{1}}\q'_{B_{2}}\ket_{\s}|\q'_{E_{x}}\q'_{E_{y}}\ket_{-\s}$. 
As before under a lattice step translation 
this state does not change its spatial 
distribution but $\s\ra -\s$ without any phase factor:
$|\q'_{A_{1}}\q'_{B_{2}}\ket_{\s}|\q'_{E_{x}}\q'_{E_{y}}\ket_{-\s}
\leftrightarrow
|\q'_{E_{x}}\q'_{E_{y}}\ket_{\s}|\q'_{A_{1}}\q'_{B_{2}}\ket_{-\s}$.

Now we use these results to diagonalize $n_{{\mathbf r}=0}$ 
on the whole set ${\cal S}_{hf}$ (we could have done that directly by 
diagonalizing $6 \times 6$ matrices but we wanted to show the general 
method). 
The eigenstate  of $n_{{\mathbf r}=0}$ with nonvanishing eigenvalue always
belongs to  $A_{1}$.
The matrix $n_{{\mathbf r}}$
has  eigenvalues $3/8$ and (5 times) $0$, as predicted by Eq.(\ref{det}). 
For ${\mathbf r}=0$ the eigenvector of occupation $3/8$ is
$|\varphi^{(0)}_{1}\rangle=
\frac{1}{\sqrt{3}}|\q''_{A_{1}}\ket+\sqrt{\frac{2}{3}}|\q'_{A_{1}}\ket$. 
The other $A_{1}$ 
eigenstate of $n_{{\mathbf r}=0}$ has 0 eigenvalue and reads:
$|\varphi^{(0)}_{2}\rangle=
\sqrt{\frac{2}{3}}|\q''_{A_{1}}\ket-
\frac{1}{\sqrt{3}}|\q'_{A_{1}}\ket$. 
The other eigenvectors, whose symmetry differs from $A_{1}$, are
$|\vf^{(0)}_{3}\ket=|\q'_{B_{2}}\ket$,
$|\vf^{(0)}_{4}\ket =|\q''_{B_{1}}\ket$, 
$|\vf^{(0)}_{5}\ket =|\q'_{E_{x}}\ket$ and 
$|\vf^{(0)}_{6}\ket=|\q'_{E_{y}}\ket$.
One finds\cite{jop2001} that the transfer matrices $T_{l}$ of 
Eq.(\ref{transferT}) such that 
$|\varphi^{(\hat{e}_{l})}_{I}\rangle
\equiv \sum_{J}|\varphi^{(0)}_{J}\rangle T_{l_{JI}}$,
are:
\begin{equation}
T_{x}=\left[\begin{array}{rrrrrr}
0 & 0 & 0 & -\frac{1}{\sqrt{3}} & i\sqrt{\frac{2}{3}} & 0 \\
0 & 0 & 0 & -\sqrt{\frac{2}{3}} & -\frac{i}{\sqrt{3}} & 0 \\
0 & 0 & 0 & 0 & 0 & -i \\
-\frac{1}{\sqrt{3}} & -\sqrt{\frac{2}{3}} & 0 & 0 & 0 & 0 \\
i\sqrt{\frac{2}{3}} & -\frac{i}{\sqrt{3}} & 0 & 0 & 0 & 0 \\
0 & 0 & -i & 0 & 0 & 0
\end{array}\right]\;,\;\;\;\;\;\;
T_{y}=\left[\begin{array}{rrrrrr}
0 & 0 & 0 & \frac{1}{\sqrt{3}} & 0 & -i\sqrt{\frac{2}{3}} \\
0 & 0 & 0 & \sqrt{\frac{2}{3}} & 0 &  \frac{i}{\sqrt{3}}\\
0 & 0 & 0 & 0 & i & 0 \\
\frac{1}{\sqrt{3}} & \sqrt{\frac{2}{3}} & 0 & 0 & 0 & 0 \\ 
0 & 0 & i & 0 & 0 & 0 \\
-i\sqrt{\frac{2}{3}} &  \frac{i}{\sqrt{3}} & 0 & 0 & 0 & 0
\end{array}\right].
\label{transfer}
\end{equation}
The reason why this choice of the basis set is clever is now 
apparent. The local basis at any site $r$ splits into the subsets
${\cal S}_{a}=\{|\vf^{(0)}_{1}\rangle,|\vf^{(0)}_{2}\rangle,
|\vf^{(0)}_{3}\rangle\}$, and
${\cal S}_{b}=\{|\vf^{(0)}_{4}\rangle,|\vf^{(0)}_{5}\rangle,
|\vf^{(0)}_{6}\rangle\}$;
a shift by a 
lattice step sends members of ${\cal S}_{a}$ into linear combinations 
of the members of  ${\cal S}_{b}$, and conversely.

Consider the 6-body eigenstate of $K$
\begin{equation}    
   |\F_{AF}\ket_{\s}=
   |\vf^{(0)}_{1}\vf^{(0)}_{2}\vf^{(0)}_{3}\rangle_{\s}
   |\vf^{(0)}_{4}\vf^{(0)}_{5}\vf^{(0)}_{6}\rangle_{-\s} .
\end{equation}
In this state there is partial occupation of 
site ${\mathbf r}=0$ with spin $\s$, but no double occupation. It turns out that a 
shift by a lattice step produces the transformation
\begin{equation}
  |\F_{AF}\ket_{\s}   \longleftrightarrow - |\F_{AF}\ket_{-\s}
\label{giochino}
\end{equation}
that is, a lattice step is equivalent to a spin flip, a feature that 
we call  {\em 
antiferromagnetic property}. Since the spin-flipped state is 
also free of double occupation, $|\F_{AF}\ket_{\s}$ is a $W=0$ 
eigenstate. A ground state 
which is a single determinant is a quite unusual property for an 
interacting model like this. 

Note that $|\vf^{(0)}_{1}\vf^{(0)}_{2}\ket$ is 
equivalent to $|\q''_{A_{1}}\q'_{A_{1}}\ket$, because this is just a 
unitary transformation  of the $A_{1}$ wave functions; so
 $|\F_{AF}\ket_{\s}$ can also be 
written in terms of the old  local orbitals (without any 
mix of the  local states of different irreps of ${\mathbf G}$): 
\begin{equation}
|\F_{AF}\ket_{\s}=|\q''_{A_{1}}\q'_{A_{1}}\q'_{B_{2}}\rangle_{\s}
|\q''_{B_{1}}\q'_{E_{x}}\q'_{E_{y}}\rangle_{-\s}.
\label{perirrep}
\end{equation}
This form of the ground state lends itself to be generalised (see below).
For $N_{\L}>4$, ${\mathbf k}$ vectors arise that do not possess any special symmetry, 
the vectors $R_{i}{\mathbf k}$ are all 
different for all $R_{i} \in C_{4v}$, and we get  eight-dimensional 
irreps of ${\mathbf G}$.  Recalling that $|{\cal S}_{hf}|=2 N_{\L}-2$,
one finds that
${\cal S}_{hf}$ contains 
$N_{e}=\frac{1}{2}(\frac{N_{\L}}{2}-2)$ irreps of dimension 8, one of 
dimension 4 and one of dimension 2 if $N_{\L}/2$ is even and $N_{o}
=\frac{1}{2}(\frac{N_{\L}}{2}-1)$ irreps of dimension 8 and one of 
dimension 2 if $N_{\L}/2$ is odd.

To extend the theory to general $N_{\L}$, we note that these ${\mathbf k}$ vectors, 
since   $R_{i}{\mathbf k}$ are all 
different, are  a basis of  the regular representation of $C_{4v}$. 
Thus, by the Burnside theorem, each of them breaks 
into $A_{1}\oplus A_{2}\oplus B_{1}\oplus B_{2}\oplus E\oplus E$;  
diagonalizing $n_{{\mathbf r}=0}$ and the point group characters on the 
basis of the $m$-th eight-dimensional irrep of ${\mathbf G}$ one gets 
one-body states  $|\q_{I}^{[m]}\ket$, where $I$ stands for the 
$C_{4v}$ irrep label, $I$=$A_{1}$, $A_{2}$, $B_{1}$, $B_{2}$, $E_{x}$, $E_{y}$, 
$E^{\prime}_{x}$, $E^{\prime}_{y}$; here  we denote 
by $E^{\prime}$ the second occourrence of the irrep $E$.
The ground state wave function in ${\cal H}$ for the half filled case is a 
generalized version of Eq.(\ref{perirrep}). For even $N_{\L}/2$, we 
have\cite{jop2001}  
\begin{equation}
|\F_{AF}\ket_{\s}\equiv|(\prod_{m=1}^{N_{e}}
\q^{[m]}_{A_{1}}\q^{[m]}_{B_{2}}\q^{[m]}_{E_{x}}\q^{[m]}_{E_{y}})
\q'_{A_{1}}\q'_{B_{2}}\q''_{A_{1}}
\ket_{\s}
|(\prod_{m=1}^{N_{e}}\q^{[m]}_{A_{2}}\q^{[m]}_{B_{1}}\q^{[m]}_{E'_{x}}
\q^{[m]}_{E'_{y}})\q'_{E_{x}}\q'_{E_{y}}\q''_{B_{1}}\ket_{-\s},
\label{detaf}
\end{equation}
with $\s=\ua,\da$. For odd $N_{\L}/2$, 
$|\F_{AF}\ket_{\s}$ is  similar  but the maximum  $m$ is   $N_{o}$
and the $|\q'\ket$ states do not occour. $|\F_{AF}\ket_{\s}$ is a $W=0$ 
state, transforms into 
$-|\F_{AF}\ket_{-\s}$ for 
each lattice step translation and manifestly shows an 
antiferromagnetic order ({\em antiferromagnetic property}). 
Since $W_{{\mathrm Hubbard}}$ is 
a positive semidefinite operator $|\F_{AF}\ket_{\s}$ is actually a ground state. 
In the basis 
of local states these are the only two determinantal states 
($\s=\ua,\da$) with the above 
properties. 

A few further remarks about $|\F_{AF}\ket_{\s}$ are in order.  
(1) Introducing the projection operator 
$P_{S}$  on the spin $S$ subspace, one finds that  
$P_{S}|\F_{AF}\ket_{\s}\equiv|\F^{S}_{AF}\ket_{\s}\neq 0\; , \forall 
S=0,\ldots,N_{\L}-1$. 
Then, 
$_{\s}\bra \F_{AF}|W_{{\mathrm 
Hubbard}}|\F_{AF}\ket_{\s}=\sum_{S=1}^{N_{\L}-1}\, 
_{\s}\bra\F^{S}_{AF}|W_{{\mathrm Hubbard}}|\F^{S}_{AF}
\ket_{\s}=0$, and this implies that there is at least one ground state 
of $W_{{\mathrm Hubbard}}$ in  ${\cal H}$ for each $S$.
The actual ground state of $H_{{\mathrm Hubbard}}$ at weak coupling 
is the singlet $|\F^{0}_{AF}\ket_{\s}$.  (2) 
The {\em existence} of this singlet $W=0$ ground state 
is a direct consequence of the Lieb theorem\cite{lieb2th}. Indeed 
the maximum spin state $|\F^{N_{\L}-1}_{AF}\ket_{\s}$ is trivially in the kernel
of $W_{{\mathrm Hubbard}}$; since the 
ground state must be a singlet it should be an eigenvector of 
$W_{{\mathrm Hubbard}}$ 
with vanishing eigenvalue.  (3) The above results and 
Lieb's theorem imply that 
higher order effects split the ground state multiplet of $H_{{\mathrm Hubbard}}$
and the singlet is lowest.  4) The 
Lieb  theorem makes no assumptions concerning the lattice 
structure; adding the ingredient of the $\mathbf{G}$ symmetry we are able 
to explicitly display the wave function at weak coupling.

Using the explicit form of $P_{S=0}$ one finds that 
$P_{S=0}|\F_{AF}\ket_{\s}=-P_{S=0}|\F_{AF}\ket_{-\s}$. This 
identity allows us to study how the singlet component transforms 
under translations, reflections and rotations. In particular the {\em 
antiferromagnetic property} tells us that the total momentum is 
$K_{tot}=(0,0)$. To make contact with 
Ref.\cite{moreodag} we have also determined how it transforms under the $C_{4v}$ 
operations with respect to the center of an arbitrary placquette. 
It turns out\cite{jop2001} that it is even under reflections and transforms as an 
$s$ wave if $N_{\L}/2$ is even and as a $d$ wave if $N_{\L}/2$ is odd. 

In the next Section we use these results, togheter with the 
non-perturbative canonical transformation, to study the doped $4\times 4$ 
lattice at half filling. Since the non-interacting ground state at 
half filling is now well known and unambiguously defined the 
expansion (\ref{psi0}) can be performed in a unique way.}

\subsection{Pairing in the Doped Hubbard Antiferromagnet}
\label{dopedaf}

{\small
Let us represent the $4\times 4$ lattice as in Figure \ref{dynsym} (A). 
\begin{figure}[H]
\begin{center}
	\epsfig{figure=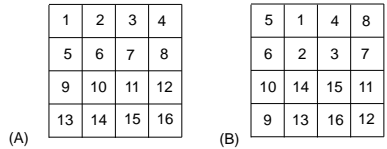,width=5cm}
	\caption{\footnotesize{
(A) The $4\times 4$ lattice. The numbers 
    in the squares enumerate the sites. (B) The $4\times 4$ lattice 
    after the {\it dynamical} symmetry operation. }}
    \label{dynsym}
\end{center} 
\end{figure}

We consider a 4$\times$4 lattice one-band Hubbard 
Model and periodic boundary conditions. 
The one particle energy spectrum of the kinetic term $K$  
has 5 equally spaced levels having degeneracy 1, 4, 6, 4, 1 respectively. 
The space Group $\mathbf{G}$ containing the translations and the 8 $C_{4v}$ 
operations is not enough to explain the degeneracy 6. Indeed the 
largest dimension of the irreps of the space Group for this 4$\times$4 
lattice is 4.  Nevertheless the  4$\times$4 lattice is very special, 
since among the symmetry operations that preserve $K$ there are some 
which are not contained in $\mathbf{G}$ and preserve $W_{\mathrm{Hubbard}}$ 
too. This fact implies that $H_{\mathrm{Hubbard}}$ is invariant under 
a new and largest symmetry Group; let us call it ${\cal G}$. 
As observed by previous authors\cite{bonca}\cite{poilb} the 
4$\times$4 lattice can be mapped in the $2\times 2\times 2\times 2$ 
hypercube since each pair of next to nearest neighbor sites 
has two nearest neighbor sites in common. Due to the importance of the 
symmetry in our configuration interaction mechanism, below  
we explicitly build a new symmetry operations and derive the 
character table of ${\cal G}$\cite{epj01}. }

\subsubsection{Symmetry Group of the $4\times 4$ lattice}

{\small 
Periodic boundary conditions are assumed and  for 
example, the nearest neighbours of 1 are 2, 5, 4 and 13. Rotating 
the plaquettes 1, 2, 5, 6 and 11 ,12 ,15, 16 clockwise and the other two 
counterclockwise  by 90 degrees we obtain the effect of the 
{\it dynamical} symmetry, that we call $d$, see Figure \ref{dynsym} (B). 
This transformation preserves nearest neighbours (and so, each order
of neighbours) but  is not an isometry, and for example the distance 
between 1 and 3 changes.  Thus, this symmetry operation $d$ is a new,
dynamical symmetry. Including  $d$ and closing the multiplication table
we obtain the Group ${\cal G}$  with 384 elements in 
20 classes (like $\mathbf{G}$). 
The complete  Character  Table of ${\cal G}$  is shown in the Table 
\ref{dynchatable}. 
\begin{table}[tbp]
    \centering
    \caption{{\footnotesize Character Table of the {\em Optimal Group} ${\cal G}$ of the 
$4 \times 4$ model.}}
    \vspace{0.5cm}
   \begin{tabular}{|c|c|c|c|c|c|c|c|c|c|c|c|c|c|c|c|c|c|c|c|c|}
	\hline
	${\cal G}$&${\cal C}_{1}$&${\cal C}_{2}$&${\cal C}_{3}$&${\cal C}_{4}$&
	${\cal C}_{5}$&${\cal C}_{6}$&${\cal C}_{7}$
	&${\cal C}_{8}$&${\cal C}_{9}$&${\cal C}_{10}$&${\cal C}_{11}$&${\cal C}_{12}$
	&${\cal C}_{13}$&${\cal C}_{14}$&${\cal C}_{15}$
	&${\cal C}_{16}$&${\cal C}_{17}$&${\cal C}_{18}$&${\cal C}_{19}$&${\cal C}_{20}$\\
	\hline
        $A_{1}$&1&1&1&1&1&1&1&1&1&1&1&1&1&1&1&1&1&1&1&1\\
	\hline
        $\Tilde A_{1}$&1&1&1&1&1&1&-1&-1&1&-1&-1&-1&-1&1&1&1&1&1&-1&-1\\
	\hline
	$B_{2}$&1&1&-1&-1&1&1&1&1&1&-1&-1&1&-1&-1&1&1&-1&-1&1&-1\\
	\hline
	$\Tilde B_{2}$&1&1&-1&-1&1&1&-1&-1&1&1&1&-1&1&-1&1&1&-1&-1&-1&1\\
	\hline
	$\G_{1}$&2&2&-2&-2&2&2&0&0&2&0&0&0&0&-2&-1&-1&1&1&0&0\\
	\hline
	$\G_{2}$&2&2&2&2&2&2&0&0&2&0&0&0&0&2&-1&-1&-1&-1&0&0\\
	\hline
	$\S_{1}$&3&3&3&3&3&-1&-1&-1&-1&-1&-1&-1&-1&-1&0&0&0&0&1&1\\
	\hline
	$\S_{2}$&3&3&3&3&3&-1&1&1&-1&1&1&1&1&-1&0&0&0&0&-1&-1\\
	\hline 
	$\S_{3}$&3&3&-3&-3&3&-1&-1&-1&-1&1&1&-1&1&1&0&0&0&0&1&-1\\
	\hline
	$\S_{4}$&3&3&-3&-3&3&-1&1&1&-1&-1&-1&1&-1&1&0&0&0&0&-1&1\\
	\hline
	$\L_{1}$&4&-4&-2&2&0&0&-2&2&0&-2&2&0&0&0&-1&1&-1&1&0&0\\
	\hline 
	$\L_{2}$&4&-4&-2&2&0&0&2&-2&0&2&-2&0&0&0&-1&1&-1&1&0&0\\
	\hline 
	$\L_{3}$&4&-4&2&-2&0&0&-2&2&0&2&-2&0&0&0&-1&1&1&-1&0&0\\
	\hline
	$\L_{4}$&4&-4&2&-2&0&0&2&-2&0&-2&2&0&0&0&-1&1&1&-1&0&0\\
	\hline
	$\W_{1}$&6&6&0&0&-2&-2&-2&-2&2&0&0&2&0&0&0&0&0&0&0&0\\
	\hline
	$\W_{2}$&6&6&0&0&-2&-2&2&2&2&0&0&-2&0&0&0&0&0&0&0&0\\
	\hline
	$\W_{3}$&6&6&0&0&-2&2&0&0&-2&-2&-2&0&2&0&0&0&0&0&0&0\\
	\hline
	$\W_{4}$&6&6&0&0&-2&2&0&0&-2&2&2&0&-2&0&0&0&0&0&0&0\\
	\hline
	$\P_{1}$&8&-8&-4&4&0&0&0&0&0&0&0&0&0&0&1&-1&1&-1&0&0\\
	\hline
	$\P_{2}$&8&-8&4&-4&0&0&0&0&0&0&0&0&0&0&1&-1&-1&1&0&0\\
	\hline
\end{tabular}
    \label{dynchatable}
\end{table}
As the notation suggests, the irreps $A_{1}$ and $\tilde{A}_{1}$ both 
reduce to  $A_{1}$, in $C_{4v}$, while  $B_{2}$ and $\tilde{B}_{2}$ both 
reduce to  $B_{2}$. 
In the Table \ref{4x4spectrum} we report the one-body 
eigenvalues of for $t=-1$,
the degeneracy and the symmetry of each eigenvector. 
For pairs, the $W=0$ theorem of Section \ref{w=otheorem} 
ensures that no double occupancy is possible in the irreps 
$\tilde{A}_{1},\;B_{2},\;\G_{1},\;
\G_{2},\;\S_{1},\;\S_{2},\;\S_{3},\;\S_{4},\;\L_{2},\;\L_{3},\;\W_{1},\;
\W_{2},\;\W_{3},\;\P_{1}$ and $\P_{2}$ not represented in the 
one-body spectrum. 
\begin{table}[tbp]
    \centering
    \caption{{\footnotesize One-body spectrum for $t=-1$.}}
    \vspace{0.5cm}
\begin{tabular}{|c|c|c|}
\hline
   Energy&Irrep of ${\cal G}$ &Degeneracy \\
   \hline
   4 & $\tilde{B}_{2}$ & 1 \\
   \hline
   2 & $\L_{4}$ & 4  \\
   \hline
   0 & $\W_{4}$ & 6 \\
   \hline
   -2 & $\L_{1}$ & 4 \\
   \hline
   -4 & $A_{1}$ & 1 \\
   \hline
   \end{tabular}
\label{4x4spectrum}
\end{table}
The exact interacting ground state of the system with 4 holes is 
threefold degenerate and belongs to an irrep  of ${\cal G}$ that contains 
the irreps $A_{1}$ and $B_{1}$ of $C_{4v}$ once and two times 
respectively\cite{ijmp00}. This irrep is not contained in the one 
particle  spectrum of $K$, which does not admit degeneracy 3.
By the $W=0$ theorem, pairs belonging to it must be $W=0$ pairs. These 
$W=0$ pairs arise from the single-particle states  of $K$ with eigenvalue 
-2, and their irrep is contained in the square of the irrep of the 
single-particle level.

The $m$ states of the expansion (\ref{psi0}) are  
all the $W=0$ pairs belonging to the ground state irrep.  The 
binding energy $|\D(4)|$ can be computed analitically by truncating 
the configuration interaction expansion (\ref{psi0}) 
to the $\a$ states and also by means of 
exact diagonalization\cite{ijmp00}. The results are 
listed below for different values of the on-site interaction $U$ 
(energies are in eV):
\vspace*{0.5 cm}
\begin{center}
\noindent 
\begin{tabular}{|c|c|c|c|c|}
\hline 
  & $U=$0.1 & $U=$0.5 & $U=$0.7 & $U=$1  \\
\hline 
$\D(4)_{\mathrm{exact}}$ & -0.053 & -0.32 & -0.67 & -1.18 \\
\hline 
$\D(4)_{\mathrm{analytic}}$ & -0.078 & -1.95 & -3.83 & -7.81 \\
\hline 
\end{tabular}
\end{center}
\vspace*{0.5 cm}
\noindent
As  expected, with increasing $U$ the difference 
$|\D(4)_{\mathrm{exact}}-\D(4)_{\mathrm{analytic}}|$ increases because the 
renormalization induced by virtual electron-hole exitations becomes 
important and it is no longer a good approximation 
to consider the $\a$ states only. Nevertheless the truncated canonical 
transformation still predicts the right sign of $\D$. 

As observed in Section \ref{afgs} the above canonical 
transformation applies when two holes are added 
to a non-degenerate vacuum. To study the system at and close to half filling,
we have to use the results of Section \ref{afgs}. In the following we 
will solve the problem of two electrons added to the half filled system.

\subsubsection{$W=0$ Pairs and Quartets}
\label{12h}

{\small
Here we use the antiferromagnetic ground 
state $|\F_{AF}^{S=0}\ket$ to predict the possible symmetries 
of the doped half filled system.  
With 12 holes, in the $U \rightarrow 0$ limit, there are two in
${\cal S}_{hf}$; the first-order ground states correspond to $W=0$
pairs.  The symmetry of these $W=0$ states can be determined {\em a 
priori} from the ${\cal G}$  irreps of Table \ref{4x4spectrum}. 
Apart from the filled 
shells, two holes go to the $\Omega_{4}$ level. 
From the character of ${\cal G}$ one can derive that 
\begin{equation}
\Omega_{4}^{2}=A_{1}+\tilde{B}_{2}+\Omega_{4}+\Gamma_{1}+
\Gamma_{2}+\Sigma_{2}+\Sigma_{3}+\Omega_{1}+\Omega_{2}+\Omega_{3};
\label{quadrato}
\end{equation}
since the first 3 entries are present in Table \ref{4x4spectrum}, 
the  $W=0$  theorem ensures that $\Gamma_{1}$,
$\Gamma_{2}$, $\Sigma_{2}$, $\Sigma_{3}$, $\Omega_{1}$, $\Omega_{2}$ and
$\Omega_{3}$ pairs have no double occupation. It turns out that the 
spin and orbital symmetries are entangled, {\it i.e.} some of these pairs 
are triplet and the rest singlet.  We can see that by 
projecting the determinantal state 
$c^{\dag}_{{\mathbf k}\ua}c^{\dag}_{{\mathbf p}\da}|0\ket$ with 
${\mathbf k},\;{\mathbf p}\in 
{\cal S}_{hf}$ on the irreps not contained 
in the spectrum.  One obtaines singlet $W=0$ pairs for 
$\G_{1},\;\G_{2},\;\S_{2},\;\W_{1}$. 
Other irreps yield $W=0$ triplet pairs. They are the three times degenerate 
irrep $\S_{3}$ and the two sixfold set of $\W_{2}$ and $\W_{3}$ 
symmetry. 
The above {\em a priori} argument hardly applies to the symmetries of 
$W=0$ quartets, because $\Omega_{4}^{4}$ contains almost every 
symmetry and we do not know any $W=0$-like theorem for quartets. 
However, we can still build the projection operators by Mathematica; 
we can project the 225 quartets on the irreps of ${\cal G}$  
and carry on the analysis in an efficient, if not elegant, 
way\cite{epj01}. 
It was found that the singlet $W=0$ quartets are 13 as many as the 
singlet $W=0$ pairs, and belong to the same irreps
$\G_{1},\;\G_{2},\;\S_{2},\;\W_{1}$. Therefore, these are the 
possible symmetries of the first-order ground states with 14 holes. 
Exact diagonalization results\cite{fop} show that for $U/t<3$ and 
$16-2=14$ holes the ground state is sixfold degenerate, with a 
doublet of states with momentum $(\p,\p)$ and a quartet with 
momentum $(\pm\p/2,\pm\p/2)$. It can be shown\cite{epj01} that 
the computed ground 
state corresponds to an $\W_{1}$ {\it electron} pair over the half 
filled system. For $U/t>3$ and the same number of 
holes a level crossing takes place: 
the ground state is threefold degenerate and contains a 
state with momentum $(0,0)$ and a doublet with momentum $(\p,0)$ 
and $(0,\p)$. The computed ground state must be assigned to a $\S_{2}$ 
{\it electron} pair over the half filled system\cite{epj01}. 
In both cases, the symmetry of the 
ground state corresponds to a $W=0$ pair. }

\subsubsection{Pairing mechanism}
\label{mechanism}

{\small 
We consider the ground state of the $4\times 4$ model with 14 holes; aside 
from the 10 holes in the inner $A_{1}$ and $\Lambda_{1}$ shells (see 
Table \ref{4x4spectrum}) the  outer $\W_{4}$ shell contains 4 holes in a $W=0$ 
quartet. We are in position to show that pairing 
between two {\em electrons} added to the  antiferromagnetic 16-holes ground 
state (half filling) comes out. 

We recall from Section \ref{12h} that, by  comparing with exact 
diagonalization results\cite{fop}, the ground state is assigned 
to  $\W_{1}$ at weak coupling and to $\S_{2}$ at a stronger coupling. 
We must preliminarily verify that  Group Theory does not forbid  
obtaining these symmetries by creating  $W=0$ electron pairs 
from the antiferromagnetic state $|\F_{AF}^{S=0}\ket$. 
This is the same as annihilating 
hole pairs. Since the state $|\F_{AF}^{S=0}\ket$ is a total symmetric 
singlet with vanishing total momentum, the labels of the 
quartets will be the same of 
the annihilated hole pairs. This operation can be 
done by hand, or with the help 
of Mathematica, and the answer to the preliminary question is 
adfirmative for $\W_{1}$ and $\S_{2}$, but not for all pairs. 
One obtains 24 $W=0$ quartets 
out of the 28 states since 
the annihilation of $\G_{1}$ and 
$\G_{2}$ $W=0$ pairs gives identically zero.
 
Next,  we have to use the antiferromagnetic ground 
state $|\F_{AF}^{S=0}\ket$ as the non-interacting ground state 
of the configuration interaction expansion (\ref{psi0}). 
In small clusters like the $4\times 4$ one the one-body states are 
widely separated and the intra-shell interaction is much more 
important than the inter-shell one; therefore, we consider only the $m$ 
states made removing two holes in ${\cal S}_{hf}$ from $|\F_{AF}^{S=0}\ket$, 
neglecting the high-lying unoccupied orbitals\cite{epj2000}. 
The explicit calculations\cite{epj01} shown that the effective interaction is 
attractive for both $\W_{1}$ and $\S_{2}$ $W=0$ pairs leading to a 
bound state with binding energy $\D_{\W_{1}}=-61.9$ meV and 
$\D_{\S_{2}}=-60.7$ meV for $U=-t=1$ eV. Therefore, the true weak coupling 
ground state is a $\W_{1}$ $W=0$ pair over the antiferromagnetic 
state. This result agree with exact diagonalization data\cite{fop}. }

\section{Conclusions}
\label{conclusions}

{\small				
We have presented the following evidence that the $W=0$ pairs are 
the quasi-particles that, 
once dressed, play the r\^{o}le of Cooper pairs: 1) as two-body 
states they do not feel the large on-site repulsion, that would 
come in first-order perturbation in any theory of pairing 
with any other kind of pairs. 2) The indirect interaction with the 
background particles gives attraction, and bound states with 
physically appealing binding energies. 3) The same results are also 
borne out by exact diagonalisation in finite clusters, if and only 
if they have the correct symmetry and filling to give raise to $W=0$  
pairs. 4) Both in clusters and in the plane, superconducting flux 
quantisation results from the symmetry properties of $W=0$ pairs.
  
The setup of our theory of  the effective interaction $W_{{\mathrm eff}}$ between two
holes (or electrons) is quite general; although we developed it in 
detail for Hubbard Models, we can readily  include phonons and other ingredients. 
In principle we can obtain $W_{{\mathrm eff}}$ by our canonical transformation, including 
systematically  any kind of virtual intermediate
states.  The closed-form analytic expression of $W_{{\mathrm eff}}$ 
we obtained includes  4-body virtual states. This describes repeated 
exchange of an electron-hole pair to all orders. One can envisage the 
pairing mechanism by spin-flip exchange diagrams that are enhanced by  
the $C_{4v}$ symmetry. 
The argument does not depend in 
any way on perturbation  theory, and the equations retain their form, 
with renormalized parameters, at all orders. The previous 
exact-diagonalisation results of cluster calculations are 
special cases. The resulting integral equation (\ref{equazione}), 
with the effective interaction (\ref{weffective}), is
{\it valid for the full plane}. Since an analytic treatment is 
prohibitive, we resort to a numerical treatment. 
We find that in the three-band Hubbard Model, 
$^{1}A_{2}$ pairs are more tightly bound close to half filling, but $^{1}B_{2}$
pairs are favored when the filling increases. We remind 
here that these symmetry labels are not absolute, but depend on the 
choice of a gauge convention. 
We get attraction and pairing at
all fillings we have considered (above half filling), but the binding energy of
the $^{1}A_{2}$ pairs drops by 
orders of magnitude as the filling increases; thus, there is no chance 
of superconducting flux quantization too far from half filling. So, 
we do not predict that superconductivity occurs outside some range of 
hole concentration. However, pairing is still there, even for large doping. 
The three-band Hubbard Model might be too idealized to allow a detailed
comparison with experiments; however we stress that  the approach presented is far
more general than the model we are using, and can be applied to more realistic
Hamiltonians. 

The above  non-perturbative canonical transformation allows to study pairing 
fluctuations when the non-interacting ground state is unique 
(non-degenerate). Nevertheless, at exactly half filling, the ground 
state is highly degenerate and the configuration interaction 
expansion of Section \ref{canonical} is not well defined any  more.  
To overcome the problem, we have 
removed the degeneracy in first order and with the help of the Lieb's theorem 
we have found the exact  ground state of the half filled one-band Hubbard 
Model at weak  coupling.  
We have shown that its 
symmetry quantum numbers are the same as in the strong coupling case. This 
result was estabilished for square lattices of arbitrary size. 

Once the degeneracy is removed the non-perturbative canonical 
transformation can be applied and the interesting question of what 
happens by doping the Hubbard antiferromagnet can be explored 
analytically. Therefore, we test the $W=0$ pairing mechanism within 
the one-band Hubbard Model with periodic boundary conditions 
using exact  numerical data on the 
$4\times 4$ square lattice\cite{fop}. In particular, several workers have 
recognised that those data could be qualitatively understood by a weak 
coupling analysis\cite{fridman}\cite{galan}; however the evidence of pairing 
was not seen and the symmetry 
analysis of the system was not complete. Furthermore, the role of 
$W=0$ states had not been discovered and the reason of the apparent 
success of the weak coupling analysis was not clarified. In our 
approach, the 
criteria exploited in the fully 
symmetric clusters of the CuO$_{2}$ plane are used to unambiguosly diagnose 
pairing. One more 
time the effective attractive interaction comes out from a spin flip 
diagram. This result lends further support to the general approach 
which predicts the existence of pairing fluctuations for systems of 
arbitrary size. 

On the other hand, the above results also prove that important 
ingredients are still missing and must be included. The $4\times 4$ 
model shows evidence of bound pairs of non-vanishing momentum, in 
degenerate representations. This opens up the possibility of charge 
inhomogeneities and Jahn-Teller distorsions, that remain to be 
explored. }

\end{document}